# Self-Supervised Elimination of Non-Independent Noise in Hyperspectral Imaging

Guangrui Ding[1,4], Chang Liu[2,4], Jiaze Yin[1,4], Xinyan Teng[3,4], Yuying Tan[2,4], Hongjian He[1,4], Haonan Lin[1,4*], Lei Tian[1,2,4*], Ji-Xin Cheng[1,2,3,4*]

[1] Department of Electrical and Computer Engineering, Boston University, Boston, MA, USA, 02215

[2] Department of Biomedical Engineering, Boston University, Boston, MA, USA, 02215

[3] Department of Chemistry, Boston University, Boston, MA, USA, 02215

[4] Photonics Center, Boston University, Boston, MA, USA, 02215

## Abstract

Hyperspectral imaging has been widely used for spectral and spatial identification of target molecules, yet often contaminated by sophisticated noise. Current denoising methods generally rely on independent and identically distributed noise statistics, showing corrupted performance for non-independent noise removal. Here, we demonstrate Self-supervised PErmutation Noise2noise Denoising (SPEND), a deep learning denoising architecture tailor-made for removing non-independent noise from a single hyperspectral image stack. We utilize hyperspectral stimulated Raman scattering and mid-infrared photothermal microscopy as the testbeds, where the noise is spatially correlated and spectrally varied. Based on single hyperspectral images, SPEND permutates odd and even spectral frames to generate two stacks with identical noise properties, and uses the pairs for efficient self-supervised noise-to-noise training. SPEND achieved an 8-fold signal-to-noise improvement without having access to the ground truth data. SPEND enabled accurate mapping of low concentration biomolecules in both fingerprint and silent regions, demonstrating its robustness in sophisticated cellular environments.

## Introduction

Hyperspectral imaging offers rich spectral and spatial information in complex environments[1-4]. Advanced vibrational spectroscopic imaging methods, such as stimulated Raman scattering (SRS)[5] and mid-infrared photothermal (MIP)[6] microscopy, provide high spatial resolution to resolve overlapped vibrational bands in biological systems. However, for low concentration molecules, the raw hyperspectral SRS or MIP images are often heavily contaminated by noise, undermining the effectiveness of subsequent spectral segmentation or quantification. Computation-based image denoising presents a viable solution to circumvent these hardware-imposed limitations. Notably, denoising algorithms that do not rely on ground truth are particularly valuable, given the challenges associated with obtaining higher-SNR images.

Model-based hyperspectral denoising algorithms rely on the prior knowledge of the targets across the spatial and spectral domains. Harnessing the spectral feature variance difference between signal and noise, principal component analysis (PCA) combined with the 2D spatial and 1D spectral wavelet shrinkage effectively reduces noise in low-energy principal components while preserving important spectral information[7]. Additionally, spectral total variation (STV) incorporates spatial and spectral smoothing constraints to mitigate noise[8]. 3D non-local means[9] and Block-matching and 4D filtering (BM4D)[10] groups the similar pixel blocks together to enhance the feature recognition in denoising. Despite these innovations, the performance of model-based denoisers often falls short in low-SNR environments due to the complexity of physical systems and challenges in deriving closed-form expressions.

To overcome these challenges, self-supervised deep learning denoising algorithms have been developed. Leveraging the premise that adjacent pixels follow the same receptive fields, Noise2Void (N2V) and Noise2Self have been developed[11,12]. Noise2Noise (N2N), on the other hand, utilizes pairs of independent low-SNR measurements, treating them as both input and output to directly learn noise statistics and object priors[13]. Multiple variants of N2V and N2N have been tailored for 3D image stack denoising. For example, DeepInterpolation[14] voids the central frame entirely, relying on the continuity of adjacent frames for signal interpolation. DeepVID[15] further introduces N2V strategy to enhance spatial signal correlations. Techniques like Noise2Stack[16] and DeepCAD[17] construct a pair of identically distributed signal and noise stacks through odd-and-even frame splitting, thereby facilitating N2N-based denoising. These state-of-art self-supervised denoising methods have successfully pushed the limits of 3D imaging systems where noise is statistically independent with identical mean. However, neither N2N nor N2V is valid for correlated noise.

Meanwhile, advanced hyperspectral imaging systems often contend with non-independent noise, due to complex signal generation, acquisition, and amplification processes. Multiple pump-probe hyperspectral imaging systems, including aforementioned SRS and MIP, rely on the demodulation of subtle intensity variance. Heterodyne detection, facilitated by a lock-in amplifier (LIA)[18], is essential for extracting weak modulation signals. However, the delayed response of the electronic circuit in the low pass filter of LIA, coupled with mismatches of signal generation frequency and extraction frequency, causes information leaking between two adjacent pixels. Besides, perturbation of the sample's local environment leads to a correlation between adjacent measurements. Additionally, spectral absorption variations among samples and inhomogeneous input energy at different wavenumbers contribute to spectral varied noise.

To address these issues, we introduce Self-permutation Noise2Noise Denoising (SPEND), a self-supervised learning-based denoiser that operates without explicit noise modeling. SPEND harnesses a stack permutation strategy, where the permutation direction is determined to break the correlation of the noise. After selecting the appropriate axis, the raw stack is split into two substacks, which are then recombined in alternating sequences (odd + even, even + odd). This configuration treats adjacent frames as two independent measurements of the same sample. Our approach not only significantly enhances the signal to noise ratio by approximately eightfold but also allows broad applicability across various spectral unmixing tools including Least Absolute Shrinkage and Selection Operator (LASSO)[19], Multivariate Curve Resolution (MCR)[20], and Phasor Analysis[21]. SPEND can provide high fidelity chemical analysis in both fingerprint region and silent region.

## Results

**Noise spatial correlation and spectral variation in a hyperspectral SRS system.**

Our study begins with an elucidation of the process for acquiring SRS images. In our SRS system, two pulse trains, pump, and Stokes, are collinearly focused on the sample, with the Stokes beam being modulated at 2.4 MHz. After interacting with the sample, the pump power loss occurs at the modulation frequency. After acquiring single color SRS images, we utilized spectral focusing to generate 3D hyperspectral SRS images (**Fig. 1a** and **Fig. S1**). The intensity of the modulation is extracted using a lock-in amplifier (LIA), which functions as a mixer and low-pass filter (**Fig. 1b**). Due to the complexity of LIA signal processing, noise in the LIA output is non-independent as detailed below.

We observed heterogeneous noise statistics along the two spatial axes. As depicted in **Fig. 1c,** the noise power spectral density (PSD), which is the Fourier transform of the autocorrelation function of the noise, along the fast and slow axes of the laser scanning differ significantly. The noise PSD along the fast axis shows a notable decreasing trend from lower frequencies to higher frequencies, indicative of a patterned noise behavior. This conflicts with the independent and identical noise condition, which would exhibit identical uniform PSD along both axes as demonstrated in two-photon fluorescence imaging (**Fig. S2**). To delve deeper into the spatial properties, we conducted a Pearson cross-correlation (PCC) analysis of noise among neighboring pixels (**Fig. 1d**). At a fixed time constant of 70% of pixel dwell time, increasing the scanning speed (i.e., reducing the pixel dwell time) invariably increases the noise correlation level. As suggested by **Fig. 1e**, reducing the time constant does not remove noise correlation. In addition, PCC is also positively correlated with the laser power on sample. These observations suggest the complexity of the underlying physical processes and frustrate the efforts to experimentally eliminate the spatial noise correlation.

Besides spatial correlation, noise in SRS is spectrally heterogeneous and correlates with signal levels. For SRS spectroscopic imaging, the constant noise hypothesis in the spectral domain no longer holds. Commonly, the well-known 1/f laser noise, detector thermal noise, and Poisson shot noise[22], which are identically and independently distributed in the spatial and spectral domain, are widely studied in SRS systems. In addition to these established noises, our investigation revealed a new type of noise that deviates from uniform distribution in a spectral domain. **Fig. 1f** shows the spectral variation of the noise based on hyperspectral SRS of dimethyl sulfoxide (DMSO) solution. This spectrally varied noise is distinct from the Gaussian noise, which has a uniform distribution, or the Poisson noise[23]. Adjustment to the detector power on the photodiode allows for fitting the Poisson noise distribution. **Fig. 1g** demonstrates significant discrepancies in the statistical behavior between Raman on-off resonance noise and Poisson noise. These findings indicate that conventional transformations from Poisson to Gaussian noise are ineffective in hyperspectral SRS systems.

We attribute the spectral varied noise to the thermal effect in a stimulated Raman process[24]

(**Fig. 1h-i**). When the frequency difference of pump and probe matches the vibrational frequency of chemical bonds, photon emissions occur leading to Raman gain and loss. Meanwhile, vibrational excitation occurs and is followed by non-radiative decay. This effect heats the local environment and leads to higher thermal fluctuation. The photothermally induced noise performs differently from the white noise, featuring a peak at the Raman resonance. Together, our results highlight the complex nature of noise characteristics in these systems, underscoring the necessity for developing a new denoising framework tailored to these nuanced dynamics. **Fig. S3** and **methods** summarize the way to noise characteristic analysis.

**Self-Permutation Noise2Noise Denoiser**

To remove the spatially correlated and spectrally varied noise in a hyperspectral SRS image, we propose a Self-Permutation Noise2Noise Denoising (SPEND) framework, depicted schematically in **Fig. 2**. We first select the permutation axis, which is determined based on the noise correlation levels along different axes, represented by the fluctuation of adjacent pixels. In **Fig. 2a**, we choose $\omega$ axis as the permutation direction due to its least correlation. This critical step informs the noise estimation of the hyperspectral SRS data. Typically, less fluctuation, indicating a higher correlation with adjacent pixels, is not representative in independent measurements, thus it should be avoided while permutation. Instead, slicing along axes with higher fluctuation can break the correlation and provide more independence between the target and input image pairs, thereby maximizing the benefits of N2N. To emphasize the importance of the choice of the permutation axis, we compared the results of multiple hyperspectral SRS data collection strategies[25]. Noise correlation properties is decided by scanning modalities. For instance, in the $\omega - y - x$ scanning modality (**Fig. S4a**), significant noise correlation exists in the spectral domain, suggesting permutating along the spatial axis yields optimal results (**Fig. S4b, c**). Conversely, in the traditional $x - y - \omega$ scanning modality (**Fig. S4d**), where spatial noise is more prevalent, permutating along the $\omega$ axis preserves essential high-frequency noise features for effective denoising (**Fig. S4e, f**).

The permutation process is illustrated in **Fig. 2b**, which involves dividing the raw data stack into odd and even slices along the chosen axis. These slices are then alternately concatenated to form the input and target datasets for the training phase, illustrated in **Fig. 2c**. The concatenation sequences provide the fundamental support for N2N, i.e., independent measurements of the same field of view. Through recombination, we can utilize the entire signal and noise information within the input data and deliver an unbiased estimation of noise. For the network architecture, we employ a U-Net structure[26], detailed in **Fig. S5**. During the training phase, the two concatenated stacks generated from the permutation step are used as the input and target of the neural network. During the prediction phase (**Fig. 2d**), the input data is fed into the model in its original sequence, maintaining the continuity and integrity of the spectral and spatial information.

To demonstrate the versatility and effectiveness of SPEND, we integrate it with several published chemical unmixing methods, including Phasor, LASSO, and MCR. In **Fig. S6** we illustrate the typical process of chemical unmixing with and without reference spectra, showcasing SPEND's ability to enhance the signal-to-noise ratio (SNR) significantly. This enhancement enables more reliable chemical unmixing, providing deeper insights into biological systems. Subsequently, we use chemical unmixing to provide comprehensive compositional information from the hyperspectral SRS stacks. Here, we demonstrate the whole process and compare the results of different permutation strategies for comprehensive study, shown in **Fig. S7**. **Fig. S7a** demonstrates that spatial permutation significantly reduces noise, although it may blur some features – suggesting potential improvements through expanded training datasets or oversampling in the spatial domain during imaging. **Fig. S7b** shows the impact of permutation choice on chemical unmixing results, emphasizing the importance of appropriate axis selection. By optimizing the selection, although the spatial information is blurred, the spectral resolution is maintained, allowing us to achieve good unmixing results. **Fig. S7c** calculates the spectral error of different permutation strategies. **Fig.S7d** shows the correlation level of each axis. The correlation along the spatial axis is lower than the spectral axis. Hence, permutation along the spatial axis is preferred. Wrong permutation strategies will impair the performance of the denoiser. **Fig. S7e, f** demonstrates the improvement comparison between the spectral domain and the spatial domain.

**Validation of SPEND performance in spectral and spatial domains**

The validation of SPEND's efficacy in enhancing hyperspectral SRS image quality is crucial, particularly as this self-supervised denoiser aims to improve the sensitivity in label-free chemical imaging modalities that lack ground truth (GT) data. The signal-to-noise ratio (SNR) in SRS imaging, which is proportional to the product of the intensity of the Stokes beam and the square root of the intensity of the pump beam $SNR \propto I_{stokes}\sqrt{I_{pump}}$, serves as a fundamental indicator of image quality [27]. To quantitatively evaluate SPEND's performance, we manipulated the intensity of the Stokes beam and pixel dwell time to create a testing set. By increasing the average power of both beams we can create high SNR references. Conversely, by decreasing the power we achieve low SNR images as testing set to be denoised. After denoising, the noise can be reduced 8.5 times, while the signal remains same level. Consequently, SNR can be increased by approximately 8.5 times. We employ the Structure Similarity Index (SSIM), Fourier Ring Correlation[28] (FRC), and Fréchet distance[29] as metrics to gauge spatial and spectral performance.

The low laser power group and denoised single-color SRS images are presented in **Fig. 3a.** The performance of SPEND is benchmarked against BM4D, a golden standard denoising algorithm. Chemical unmixing was pursued (in **Fig. 3b**) using LASSO, with references shown in **Fig S8**, alongside their unmixed hyperspectral SRS counterparts for lipid, cholesterol, and protein channels. We further recorded an image of the same sample at high laser power as a reference to validate the accuracy. **Fig. S9** shows the high SNR group and its unmixing result which are used as the reference to calculate spectrum error and SSIM. The spectral error was evaluated by calculating the Fréchet distance based on the single-pixel spectrum (**Fig. 3c**, **Fig S10** and **Methods**). SPEND exhibited less spectral distortion compared to both BM4D and low power data. In **Fig. 3d-e**, SSIM and PSNR are conducted on each chemical channel evaluating the spectral and spatial performance of SPEND. SPEND shows better performance over BM4D. In **Fig. 3f-g**, a line in the ROI is highlighted. There are no discernible features in the low power and BM4D group. However, after applying SPEND, the 1 $\mu m$ diameter spot at 2 $\mu m$ position can be identified. Next, we quantified the resolution by Fourier Ring Correlation (FRC). FRC provides a measure of the consistency of the spatial frequency information. As shown in **Fig 3h**, the resolution achieved by SPEND approached 400 nm, which closely matches that of the high laser power group and significantly surpasses the around 600 nm resolution in low laser power and BM4D groups. Overall, SPEND outperforms BM4D in SNR and spatial resolution improvement.

**SPEND enables high fidelity SRS imaging in the fingerprint region**

The spectrally crowded SRS signals in the carbon-hydrogen (C-H) stretching vibration region (2800-3100 cm$^{-1}$), where strong Raman bands reside, have limited chemical specificity in a complex biological environment. Instead, fingerprint SRS offers a significant advantage in enhancing specificity by providing distinct Raman peaks for each biological component. Major chemicals such as protein, fatty acid, and cholesterol can be distinguished in this region, shown in **Fig. 4a**. However, the fingerprint region presents challenges due to the inherently weak Raman cross-sections, which tend to reduce the SNR, posing difficulties for chemical analysis.

As shown in **Fig. 4b,** applying SPEND to a hyperspectral SRS image stack significantly improves the SNR within the fingerprint window. We cross-validate with the C-H region. A single frame is displayed to illustrate the enhancement. A detailed comparison is provided in **Fig. S11,** highlighting the spectra in the regions rich in fatty acids, cholesterol, and proteins, illustrating a considerable reduction of spectral noise by SPEND. Co-plotted spectra (**Fig. S11b-d)** from these selected regions underscore the improvement of SPEND. Chemical maps generated by MCR from raw fingerprint data, SPEND group, and C-H region are shown in **Fig. 4c**. MCR can retrieve the chemical spectrum in the cell environment and enable robustness analysis by comparing the difference of input and retrieved spectrum. We first use the SSIM to compare the similarity between the unmixing result of the fingerprint region and CH region, shown in **Fig.4d**. In the raw data group, it is challenging to separate the fatty acid, cholesterol, and protein components. A lot of background noise displays signals in fatty acid and protein channels. In the box-plot SSIM result, the value of raw data is close to zero, which means non-similarity between low SNR raw fingerprint data and high C-H data. After denoising, the chemical component can be separated into each channel. The high value of SSIM demonstrates the significant similarity of each chemical map with the C-H result. However, we still observe some differences in unmixing results between fingerprint and C-H region, as pointed out by the arrows in **Fig. 4c**. The diffused cholesterol is shown as concentrated dots in C-H unmixing results. In protein channel, strong aggregation region is caused by the leakage from DNA. We believe fingerprint region provides more precise separation. C-H region suffers from the cross-talk of different chemical components. By comparing the difference between the retrieved spectrum and the pure chemical spectrum, we can evaluate the accuracy of the unmixing result. **Fig. 4e** compares the distortion in retrieved spectra from MCR in both fingerprint and C-H regions. The average value of both fingerprint and C-H region is at the same level, indicating credible unmixed chemical maps. However, the variance of the C-H region is notably higher than the fingerprint region due to the crowded chemical peaks in the C-H region, leading to increased uncertainty in chemical unmixing results. Some direct comparisons are shown in **Fig. S12**. In fingerprint, MCR maintains the peaks of each component at the same wavenumber. Although a shoulder shows up around 1600 $cm^{-1}$ which is caused by the cross-phase background information leak, there is limited impact to unmixing result due to relatively low intensity. In the C-H region, the overall spectrum shape of fatty acids and cholesterol is changed. Fatty acid peak around 2850 $cm^{-1}$ and cholesterol peak around 2860 $cm^{-1}$ have been buried in the background. In general, application of SPEND in fingerprint region addresses the weak Raman signal issues and improves chemical unmixing accuracy.

## SPEND enables high-fidelity SRS imaging of low-concentration molecules in the silent region

SRS has been extensively used for biorthogonal imaging of alkyne-tagged small molecules in the silent window[30]. For example, homopropargylglycine (HPG) as a biorthogonal precursor is used to study protein synthesis in click chemistry[31,32]. HPG exhibits a Raman peak at around 2125 $cm^{-1}$, SRS imaging of HPG at the silent window allows for probing protein dynamics and distribution in a click-free manner[33]. Despite its importance, quantitative analysis of the SRS signals has been complicated by low SNR attributable to the limited cross-section and the micromolar-level concentration of HPG. Additionally, cross-phase modulation (XPM) significantly contributes to the background noise, further complicating the isolation of pure HPG signals.

By applying SPEND, we markedly improved the fidelity of SRS images within the silent region. To facilitate quantitative chemical unmixing, the accurate reference spectra for each component, including HPG and XPM, are essential. We derived the HPG spectrum from the hyperspectral SRS of a pure HPG solution. Next, we used spectral phasor analysis to obtain other references for unmixing (**Fig. 5a-f**). Phasor analysis identified three distinct clusters, representing HPG, XPM in the cell body, and XPM in lipids. The clarity is significantly enhanced in both HPG treated and control samples, confirming the robustness of SPEND. **Fig. S13** showcases the noise reduction effect by plotting the spectrum from selected regions of interest (ROIs), confirming the suppression of frame-to-frame noise across all pixels. Using asymmetrically reweighted penalized least squares (arPLS)[34], we extract the HPG peak for each ROI (**Methods**). Without SPEND, extracting such peaks was unfeasible in such regions due to the variable XPM spectrum among samples. Surprisingly, XPM for each component within the cell was found to be different, leading to uncertainty in traditional chemical unmixing methods. The clarity required for unmixing was unachievable with raw data, underscoring SPEND's efficacy in distinguishing these components. Phasor retrieved spectrum served as references for LASSO unmixing, detailed in **Fig. S14**.

Through SPEND and unmixing, we differentiated major components in the hyperspectral SRS stack, including HPG, XPM of lipids, and cell bodies (**Fig. 5g,h**). Frame-to-frame variation analysis revealed substantial noise reduction in both HPG treated and control groups (**Fig. 5i**). In **Fig 5j**, we calculate the HPG intensity of 12 cells to confirm the stability of this method. In the HPG treated group, the intensity in the HPG channel is significantly higher than that in the control group, with a p-value of 1.58e-7, affirming the effectiveness of SPEND.

**SPEND enables high-fidelity hyperspectral MIP imaging in the fingerprint region**

Besides the SRS modality, MIP imaging[35-38] is another potent tool for revealing chemical distribution in a label-free manner. **Fig. 6** showcases the complex noise distribution in a MIP system[39] and demonstrates the improvements that SPEND brings to a hyperspectral MIP dataset. **Fig. 6a** illustrates the physical concept of MIP, where a Mid-IR beam excites the chemical bonds and arouses temperature in the sample. This heat alters the refractive index, affecting the redistribution of the visible probe beam intensity, which can be used to generate chemical maps. **Fig. S15** shows the diagram of a MIP microscope. We analyzed noise correlation in MIP images of PMMA beads in **Fig. 6b**, underlying the non-independent nature of noise distribution in MIP microscopy. **Fig. 6c** shows noise spectral variation, supporting the universality of spectrally varied noise across all pump-probe imaging modalities and underscoring the necessity of our method.

Next, we demonstrate SPEND's effectiveness on MIP images taken in the fingerprint region. In **Fig. 6d**, the SNR improvements in the hyperspectral MIP image of *C. albicans* fungal cells in the fingerprint region are displayed. After SPEND, the SNR is improved by 13 times. Three single wavenumbers from the hyperspectral MIP data stack are presented, including $952cm^{-1}$, $1070cm^{-1}$, and $1162cm^{-1}$. The first two wavenumbers correspond to the O-H and C-O bond in carbohydrates, crucial fungal cell wall components. The concentrations of O-H and C-O are different. O-H band originated from Polygalacturonase acid (PGA), commonly found in the plant cell wall[40], and is less common in the fungal cell wall. In the raw data, it is hard to distinguish individual fungal cells due to the low SNR. After SPEND, the differential distribution of O-H and C-O becomes clear. 1162cm$^{-1}$ corresponds to the C-O bond in lipids. After SPEND, several lipid droplets can be resolved. Temporal color encoding is then employed to show the three-layer structure of the cell wall (**Fig. 6e, f**). In the raw data, the low SNR obscured continuous layer structures. However, after denoising, the structure delineation appears smoother and more distinct. The spectrum is shown in **Fig. 6g**, further validating the accuracy along the spectral dimension. In summary, SPEND significantly enhances SNR in MIP imaging within the fingerprint region, enabling clearer differentiation and visualization of complex biochemical structures.

## Discussion

Denoising is essential to improve sensitivity across all kinds of microscopy. In this work, based on hyperspectral SRS microscopy and MIP microscopy, we provide an in-depth analysis of the sophisticated noise distribution. Spatially, noise correlation arises from signal leakage in the lock-in amplifier and perturbation diffusion in the sample, predominantly along the fast-scanning axis. An effective way to reduce this correlation is by increasing the pixel dwell time, though this concurrently reduces scanning speed. The heat accumulation may contribute to the correlation as well, thus corrupting the effect of longer pixel dwell time. Spectrally, the on-off resonance status induces noise variation nearly proportional to the SRS signal intensity.

To mitigate the complicated noise issues, we developed SPEND, a self-supervised framework. We demonstrate the necessity of permutation to improve the accuracy and efficiency. SPEND enhances the sensitivity by up to eight times and proves compatibility with various chemical unmixing tools, thus facilitating high-fidelity SRS and MIP imaging in both fingerprint and silent regions. In the fingerprint region, SPEND enables quantitative analysis of lipids, proteins, and cholesterol, which are hard to distinguish in the CH region. By suppressing spectral noise and enhancing the precision of chemical channel separation, SPEND refines chemical unmixing in the fingerprint region, even in the absence of high SNR reference. In the silent region, SPEND enables quantitative distribution analysis of HPG, which cannot be accomplished by single-color SRS. Furthermore, the application of SPEND to MIP microscopy demonstrates its versatility across other kinds of imaging modalities. Compared to SRS imaging modalities, MIP is more sensitive in the fingerprint region benefiting from the large infrared absorption cross section. SPEND further boosts the performance enabling better SNR of MIP images.

Beyond SRS and MIP imaging, SPEND has the potential to assist other lock-in based 3D imaging modalities that suffer from complex noise challenges. Techniques like stimulated Raman photothermal microscopy[24], Brillouin scattering microscopy[41] and transient absorption microscopy[42], which encounter similar sophisticated noise issues, could benefit from the improved SNR without necessitating additional hardware expenses. Moreover, the application of SPEND extends beyond hyperspectral imaging; it could also enhance SNR in videos, providing a convenient solution to suppress noise issues in dynamic imaging scenarios.

While SPEND demonstrates robust denoising capabilities, it is also possible to encounter misfitting problems. These challenges can be addressed by expanding the training dataset's size and complexity, which helps refine the model's accuracy and generalizability. However, achieving the independent measurements required by the N2N framework poses specific demands on data acquisition strategies in both spatial and spectral domains. To optimize the effectiveness of SPEND, we recommend adhering to the Nyquist sampling rate during data acquisition, which should be tailored according to the specific features of interest in the post-denoising analysis. Thus, a strategic approach to data acquisition is crucial to fully leverage

SPEND's denoising capabilities and ensure high-quality imaging outcomes.

## Methods

### Stimulate Raman scattering microscope

A lab-built hyperspectral stimulate Raman scattering microscope (**Fig. S1**), previously published[43], is used to perform hyperspectral SRS imaging. A femtosecond pulse laser (Insight, DeepSee+, spectra-Physics) operating at 80MHz with two synchronized beams, a tunable pump beam ranging from 680nm to 1300nm, a fixed stokes beam at 1040nm. The pump beam is tuned to 800nm for the C-H region, to 852nm for the CC triple bond in the silent region, and 890nm for the fingerprint region. The Stokes beam is modulated at 2.5MHz by an acoustic optical modulator (1205-C, Isomet) and chirped by a 15cm glass rod (SF57, Schott) before the merging of two beams. The combined two beams were chirped by five glass rods to picosecond pulse. A motorized linear stage is used to tune the time delay between the pump and Stokes pulse which corresponds to the Raman shift of chemical bonds. A 2D Galvo scanner (GVS102, Thorlabs) is used for laser scanning. The combined beam is sent to the sample through a 60X water immersion objective (NA=1.2, UPlanApo/IR, Olympus). After interacting with the sample, the beam is collected by an oil condenser (NA=1.4, U-AAC, Olympus). A photodiode (S3994-01, Hamamatsu) is used to collect signals after filtering the Stokes beam. The lock-in amplifier (UHFLI, Zurich Instrument) is used to extract high frequency signals.

### Noise spectral and spatial analysis

The evaluation of noise property was performed in **Fig. S3**. To simplify the process, we conduct all the analysis on pure chemical samples. For analyzing spectral variation, we measured standard deviation within a small area to quantify noise level. Then the average intensity of the same area was calculated to represent the signal level. We, then, plotted the relationship between the noise and signal to elucidate their dependency. Spatial correlation analysis was conducted using videos of pure chemical samples. Due to the objective drift issues, we used the average intensity to represent the signal. The individual frame was then subtracted from this average to isolate the noise distribution. Pearson cross-correlation analysis was then performed between adjacent rows or columns to assess the spatial correlation of noise.

### Training and interference

To enhance the robustness of our model, we implemented data augmentation such as flipping and rotating at 180 degrees. This was done to avoid distorting the inherent noise patterns, which are non-uniformly distributed. Such a cautious approach is essential as traditional data augmentation methods can potentially exacerbate noise issues if not aligned with the noise structures. This augmentation ensures the integrity of the augmented datasets is maintained, accurately reflecting the characteristics of the original data.

After the augmentation, the size of the training set increased fourfold. The training set is comprised of 24 stacks, with 10% for validation and 90% for training. Each stack contains 400*400 pixels, and 100 frames. We employed a 4-layer Unet architecture based on the CSBDeep framework[44]. The training was conducted on a commercial graphics processing unit (GPU, RTX 4090, Nvidia), taking 2 hours to complete. For interference, it will take 14 seconds to denoise an entire image stack.

### Chemical unmixing

In the paper, we utilized three established methods for chemical unmixing, MCR, Lasso, and phasor. The dimensions of the hyperspectral data, x, y, $\lambda$, dissected as $N_x$, $N_y$, $N_\lambda$.

For spectral phasor analysis, we interpreted the spectrum of each pixel through the discrete Fourier transform of first-order harmonics. This analysis was facilitated by scattering the pixels of the entire image across the complex plane, allowing us to identify specific clusters representing target chemical channels. The phasor was performed by the ImageJ plugin (Spechron Phasor)

For MCR and Lasso, we first reshape the 3D hyperspectral stack into a 2D matrix ($D \in R^{N_x N_y \times N_\lambda}$) by arranging the pixels in the raster order. Assume the number of interested chemical channels is $K$, a model is used to decompose the data matrix into the multiplication of concentration maps $C \in R^{N_x N_y \times K}$ and spectral profiles of pure chemicals $S \in R^{K \times N_\lambda}$:

$$D = CS^T + E \qquad (1)$$

where E is the error. MCR-ALS is an algorithm that solves the bilinear model using a constrained Alternating Least Square algorithm, which improves the interpretability of the profile in both $C$ and $S^T$. In Lasso, we add an L1-norm regularization to each row of the concentration matrix and solve the original inverse problem in a row-by-row manner through LASSO regression:

$$\widehat{C_{i,:}} = argmin_{C_{i,:} \geq 0} \left\{ \frac{1}{2} \left\| D_{i,:} - C_{i,:} S^T \right\|_2^2 + \lambda \left\| C_{i,:} \right\|_1 \right\} \qquad (2)$$

Where $C_{i,:}$ is a k-element nonnegative vector representing the $i_{th}$ row of the concentration matrix, $\widehat{C_{i,:}}$ is the output of LASSO regression, $D_{i,:}$ is the $i_{th}$ row of the data matrix, $\lambda$ is the hyperparameter that tunes the level of sparsity. MCR was implemented by a Python library, pyMCR. Lasso can be achieved by a GitHub project (github.com/buchenglab/LASSO-spectral-unmixing)

**Quantitation of spectrum distortion**

We employed the Fréchet distance of input spectrum and reference spectrum to quantify spectrum distortion in a pixel, shown in **Fig S10a**. In mathematics, Fréchet distance measures the similarity between two curves by calculating the minimum distance of each point. Let A and B be two curves represented by continuous functions $a: [0,1] \to R^n$ and $b: [0,1] \to R^n$, where $R^n$ is an n-dimensional Euclidian space. The Fréchet distance between A and B is given by $d_F(A, B) = \inf_{\alpha, \beta} \max_{t \in [0,1]} \|a(\alpha(t)) - b(\beta(t))\|$, where $\alpha$ and $\beta$ range over all continuous, non-decreasing surjective functions from [0,1] to [0,1], ensuring that both curves are traversed from start to finish. The norm || || typically represents the Euclidean distance between points on the two curves.

In the spectrum distortion part, we calculate the Fréchet distance of each pixel. **Fig. S10b,c** shows an example of calculating spectral distortion. Parallel computing was used to improve time consumption.

**Spatial resolution calibration**

We use Fourier ring correlation[28] to benchmark the resolution and by accounting for both the PSF and SNR of the image. It can calculate the normalized cross-correlation between the Fourier image pairs (Input and reference) at different spatial frequencies and find the cutoff spatial frequency as the cross-correlation reduces to 1/7.

**Asymmetrically reweighted penalized least squares (arPLS) smoothing for peak**

extraction

The arPLS is a numerical method for baseline correction. With a high tolerance of noise in the input spectrum, it performs well to remove the cross-phase background in hyperspectral SRS data. Assuming x is the input signal vector and z is the underlining background, z must keep the trend of x with smoothness, expressed by the regularized least square function:

$$R(z) = (x-z)^T(x-z) + \lambda z^T D^T Dz \tag{3}$$

where D is the difference matrix. Putting weights term in equation 3, it was modified into a penalized least square function:

$$P(z) = (x-z)^T W(x-z) + \lambda z^T D^T Dz \tag{4}$$

Pushing the partial derivative $\frac{\partial P}{\partial z^T} = 0$, the result of z can be expressed as

$$z = (W + \lambda D^T D)^{-1} Wx \tag{5}$$

The PLS algorithm changes weights iteratively, comparing each estimated baseline $z_i$ and signal $x_i$. To eliminate noise interference, in arPLS algorithms, the asymmetric weighting mechanism incorporated by a logic function:

$$w_i = \begin{cases} logistic(d_i, m, \sigma) = \dfrac{1}{1 + e^{\frac{2(d_i - (-m + 2\sigma))}{\sigma}}}, x_i > z_i \\ 1, x_i < z_i \end{cases} \tag{6}$$

where $d_i = x_i - z_i$, and $m, \sigma$ are the mean and standard deviation of the negative d region. The algorithm iterates until convergence, that is when reaches the maximum number of iterations or weights change smaller than $\frac{|w_t - w_{t+1}|}{|w_t|} < r$, where $r$ is the ratio parameter, $w_t$ and $w_{t+1}$ are weights at t and t+1 iteration.

**Sample preparation**

**OVCAR5 cells**

OVCAR5 cells were cultured in PRMI 1640 medium supplemented with 2 mM L-glutamine, 10% fetal bovine serum (FBS) (v/v), and 1% penicillin/streptomycin(P/S) (v/v). Then they cultured at 37°C in a humidified incubator with 5% $CO_2$ supply.

**U87 cancer cells**

After seeding the U87 cells in a 35mm glass-bottom dish overnight, the original culture medium was replaced with media containing PhDY-Chol. Cells were incubated within the medium that contains analogs at a concentration of $20\mu M$ for 48 hours. Cells were fixed with 10% neutral buffered formalin for 30 minutes and then washed with phosphate-buffered saline (PBS) three times.

**HPG-treated SJSA-1 cancer cells**

SJSA-1 cells were cultured in RPMI-1640 medium, supplemented with 10% (v/v) (FBS) and 1% (v/v) P/S.

The Homopropargylglycine (HPG) treated cells were incubated with methionine-deficient medium with 2mM HPG supplied in the medium for 24 h. While the HPG Ctrl group were incubated in methionine-deficient medium only.

All the cells were cultured under humidified incubator with 37℃ and 5% $CO_2$ supply. Finally, all the cells were fixed with 10% neutral-buffered formalin for 30 minutes, followed by PBS washes before microscopic imaging.

### C. albicans fungal cells

C. albicans isolates were cultured using yeast extract peptone dextrose medium. The culture was incubated overnight at 37°C with continuous shaking at 250 revolutions per minute. Following incubation, a 500-µl suspension of Candida was centrifuged and the pellet obtained was washed three times with phosphate-buffered saline (PBS) to remove any residual medium. The washed cells were then resuspended in PBS. Approximately 15 µl of the *Candida* suspension was carefully placed onto the surface of a 0.2 mm thick calcium fluoride ($CaF_2$) substrate. This setup was then covered with a 0.15-mm-thick coverslip to create a sandwiched structure, optimizing the optical path and minimizing cell movement during imaging. Photothermal imaging was conducted by detecting the signal from forward scattering.

### Acknowledgements.

The authors thank Zhongyue Guo for discussion in writing process. This work is supported by NIH grants R35GM136223, R01 EB032391, and R01 EB035429 to JXC.

### Author contributions.

G.D. built the SPEND network, conducted hyperspectral SRS experiments, and analyzed the data. H.L. provided the OVCAR-5 dataset. Y.T. provided the U87 hyperspectral SRS dataset. X.T. and H.H. prepared HPG-treated SJSA-1 cells. J.Y conducted the MIP experiment. C.L. helped data discussion. H.L., L.T., and J.C. supervised the project. G.D., H.L and J.C. co-wrote the manuscript. All authors read the manuscript.

### Data availability.

The dataset and code have been made publicly available on https://github.com/buchenglab.

# Figures and Figure Captions

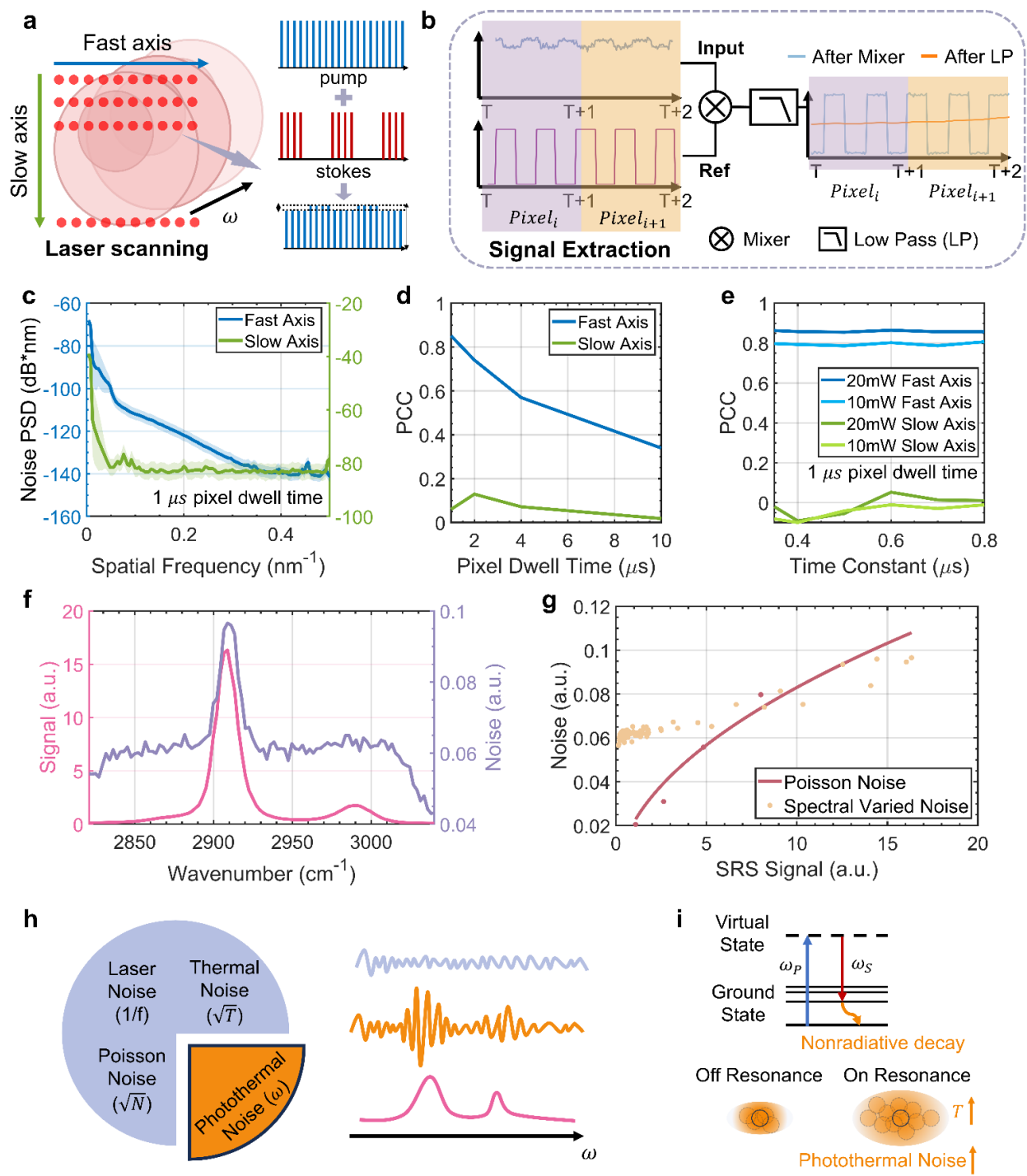

**Figure 1. Hyperspectral SRS spatial and spectral noise analysis.** (a) Illustration of hyperspectral SRS laser scanning process. (b) Schematic of the lock-in amplifier for signal extraction. LIA: Lock-in amplifier. SIG: signal input channel. REF: reference input channel. (c) Noise power spectral density (PSD) is measured along different axes. (d) Analysis of noise spatial correlation to pixel dwell time. (e) Comparison of noise spatial correlation across different axes under varying power settings. (f) Spectral varied noise. Noise and signal intensity distribution via the wavenumber. (g) Comparison of Poisson noise with spectrally

varied SRS noise. (f) Comparison of spectral varied noise induced by the photothermal effect and white noise in hyperspectral SRS system. The most widely studied white noise contains laser noise, thermal noise, and Poisson noise. (g) Physical diagram depicting the generation of photothermal noise.

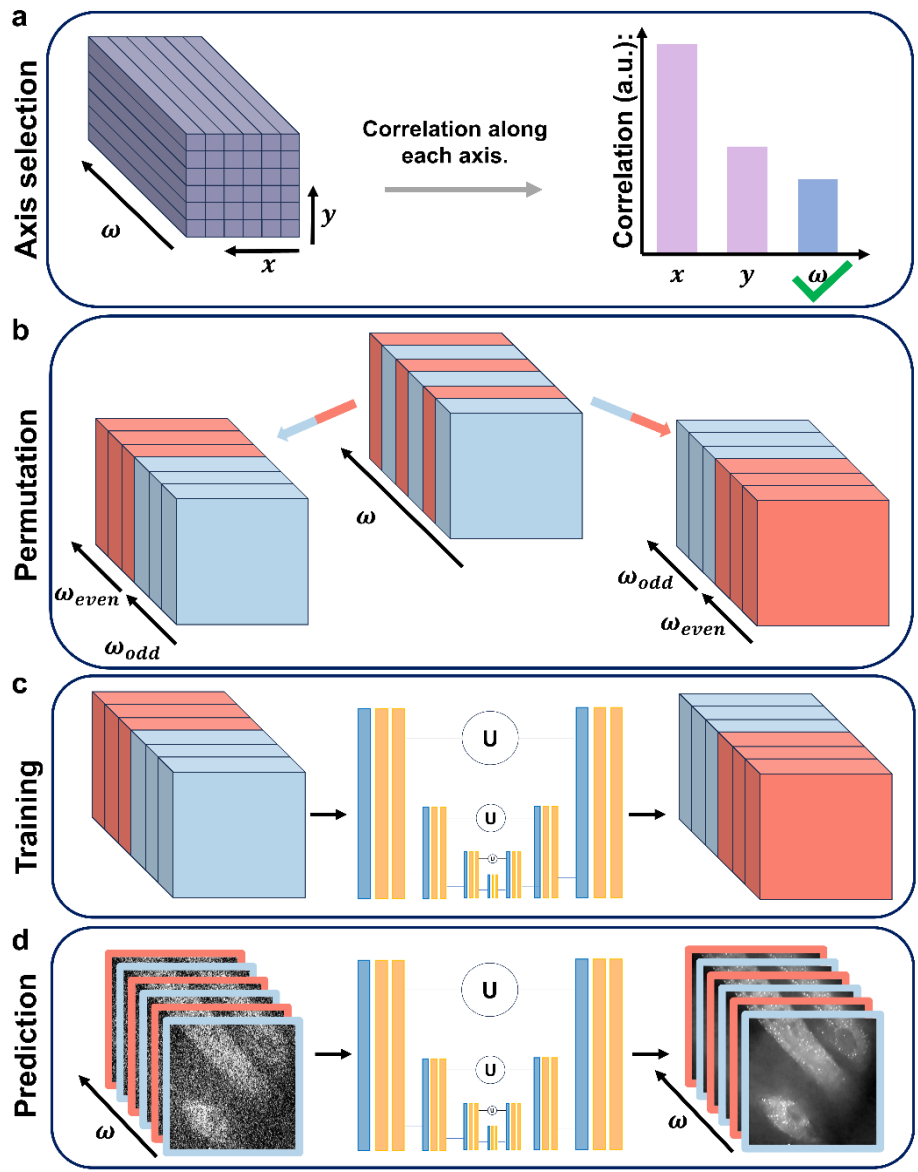

**Figure 2. SPEND workflow.** (a) Axis selection. Directions with the highest fluctuation, indicative of the lowest correlation axis, will be chosen as the permutation axis. (b) Permutation. Upon selecting the axis, the data cube is rearranged into two different sequences, which make the training pairs for the next step. (c-d) Training and prediction phase. In training, the input and result data set are generated from permutation. In prediction, the input data will maintain the original sequence without additional process. Unet is selected as a neural network model.

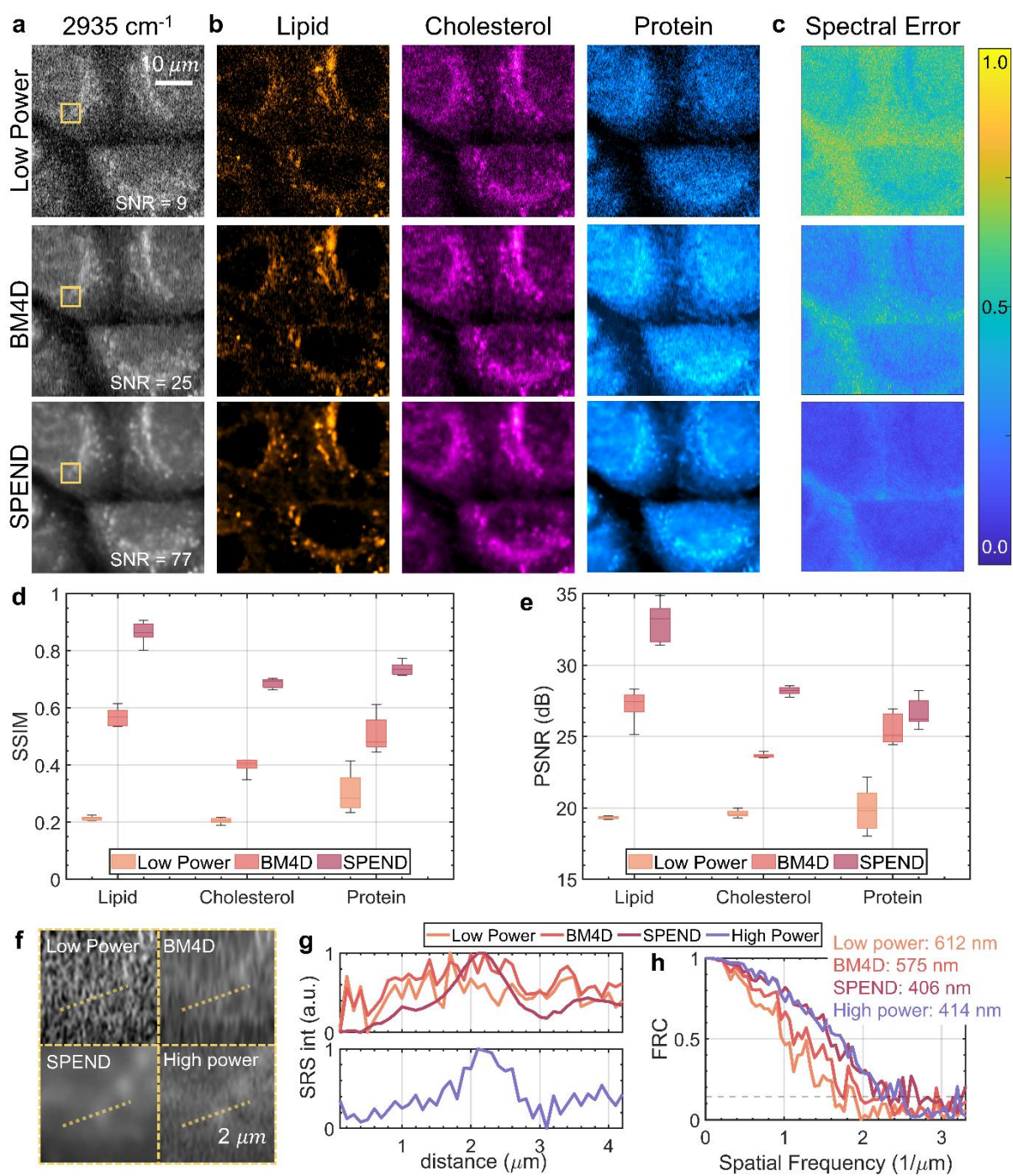

**Figure 3. Spatial and spectral fidelity analysis of SPEND.** SPEND is compared with BM4D. Using maximum power to generate high SNR reference to validate denoising result. (a) Single color SRS of OVCAR5 cancer cell taken at 2935.2 cm$^{-1}$. (b) Chemical unmixing map. The hyperspectral SRS data stacks are divided into 3 chemical channels, including Triacylglycerols (TAG), Cholesterol (CHL), and Bovine Serum Albumin (BSA). (c) Spectrum error respectively. The value indicates the difference compared with the ground truth dataset. (d) SSIM analysis of each chemical channel. (e) PSNR analysis of each chemical channel. (f) Pinpoint small ROI. (g) Dashed line intensity plot. (h) Resolution calibration by FRC. Low power: 612 nm; BM4D: 575 nm; SPEND: 406 nm; High Power: 414 nm.

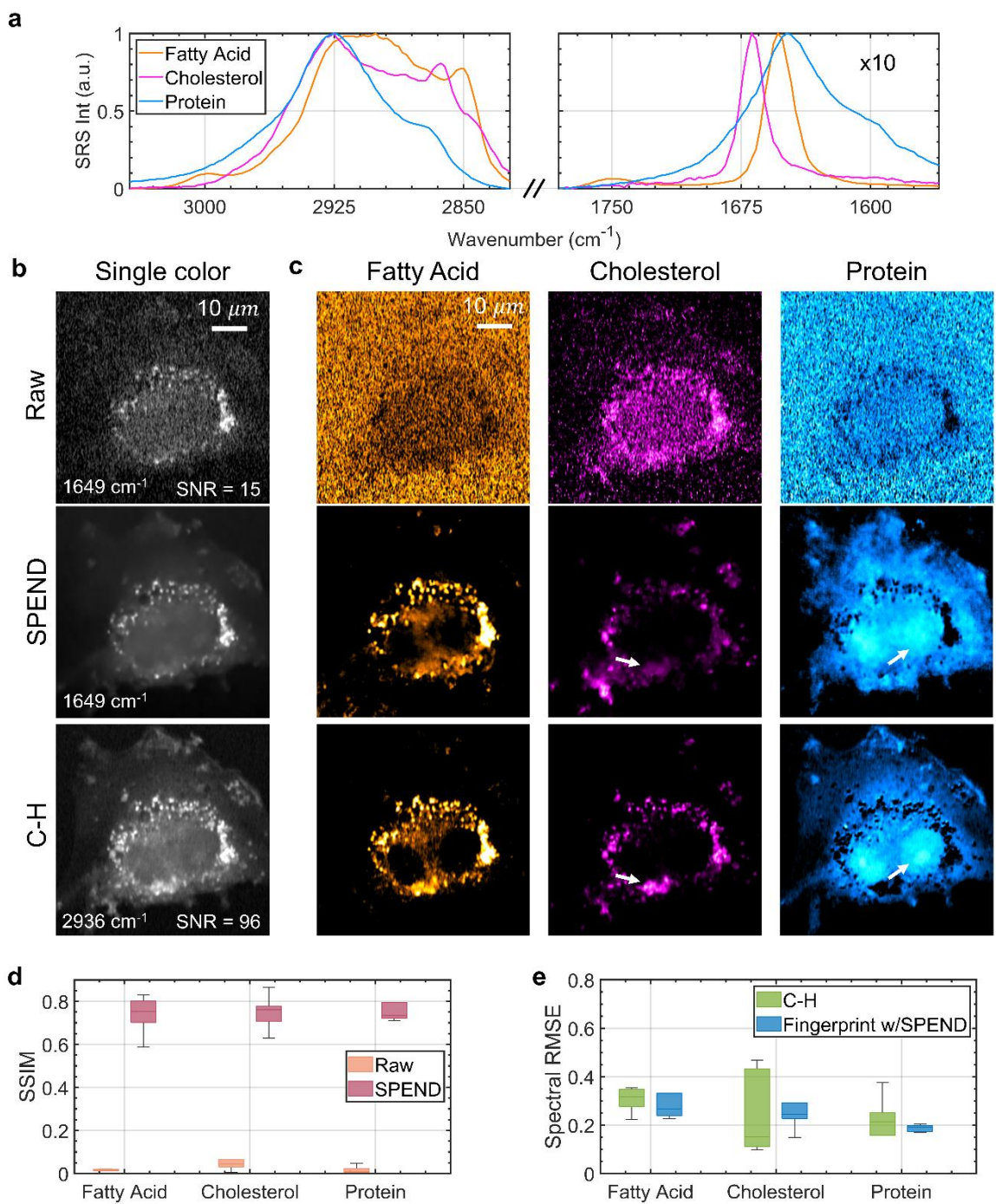

**Figure 4. SPEND enables high fidelity SRS imaging in fingerprint regions.** (a) SRS spectra of fatty acid, cholesterol, and protein in the C-H and fingerprint regions, respectively. Spectra in fingerprint region are magnified by 10 times for clarity. The references are also used for MCR unmixing. (b) Single color SRS imaging of U87 cell in fingerprint region at 1649 cm$^{-1}$ and C-H region at 2936 cm$^{-1}$. (b) Chemical map unmixing result. The SRS data was divided into 3 channels including lipid, cholesterol, and protein. We cross-validate with the unmixing result from both fingerprint and C-H regions. (d) SSIM boxplot of SPEND in denoising fingerprint hyperspectral SRS dataset. (e) RMSE between retrieved spectra after MCR and input chemical spectra from pure chemicals. Raw data was published[19].

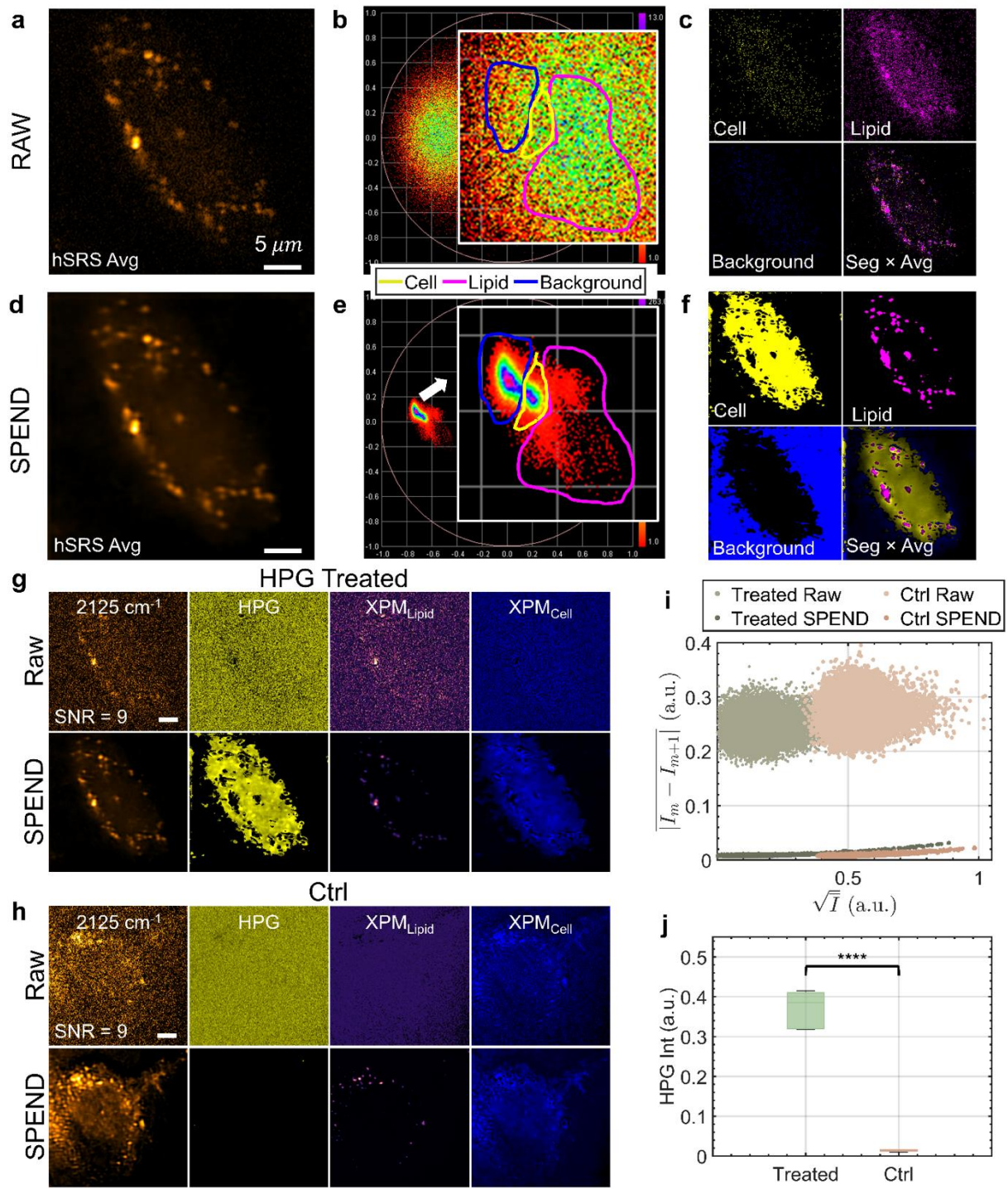

**Figure 5. Quantitative HPG mapping via hyperspectral SRS and SPEND in the silent region.** (a-c) Phasor analysis of raw HPG treated SJSA-1 cell. (d)-(f) Phasor analysis of HPG treated SJSA-1 cell after SPEND denoising. Three clusters show up in spectral phasor after denoising. (g) HPG treated group LASSO unmixing result. (h) Control group LASSO unmixing result. (i) Frame-to-frame noise reduction analysis. (j) Statistical analysis of HPG intensity in treated and control groups. p=1.58e-7<0.005.

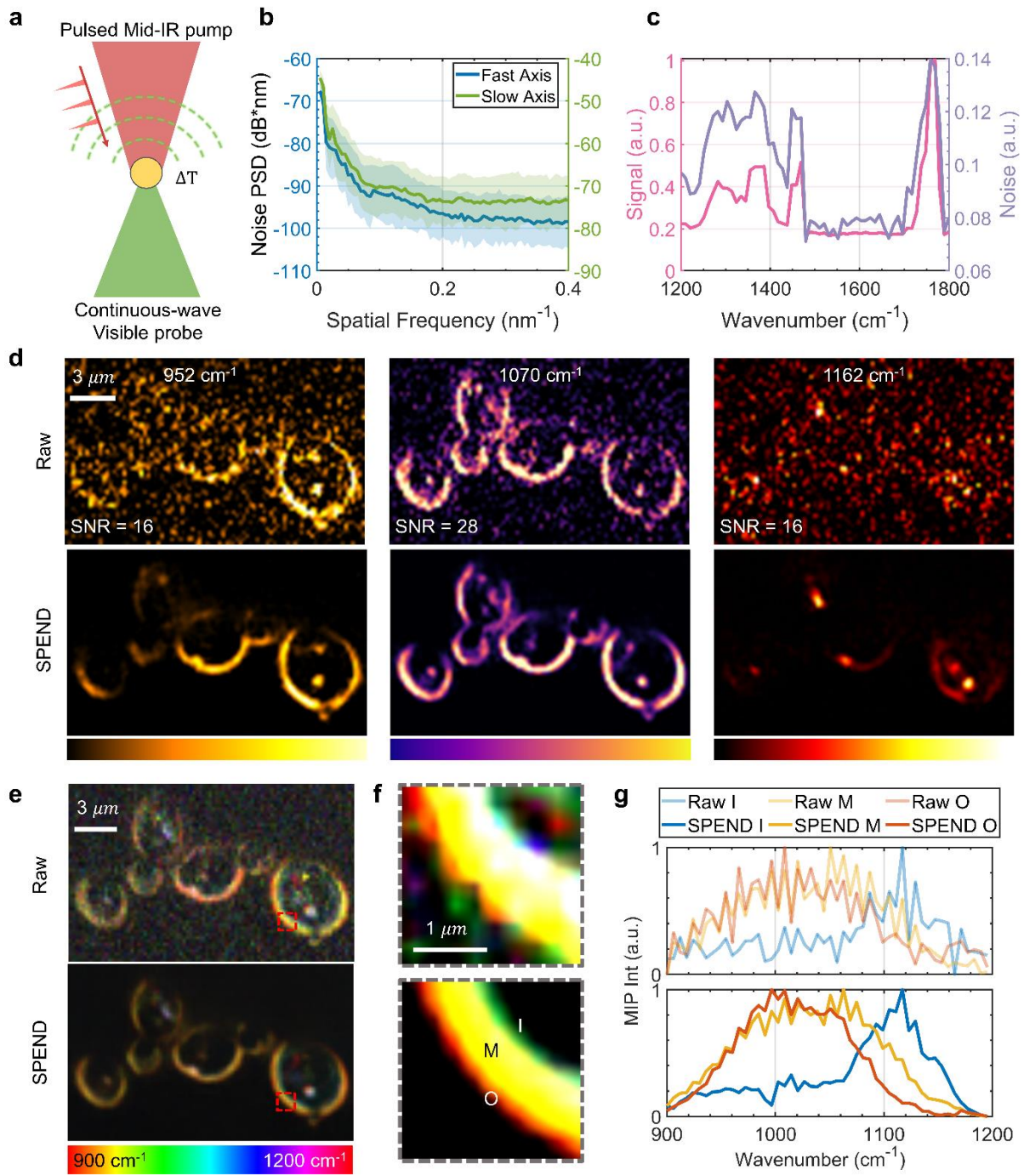

**Figure 6. SPEND performance for hyperspectral MIP microscopy.** (a) MIP diagram. Pulsed mid-infrared pulses are used to excite molecules to the vibrational state. Nonradiative decay in the process leads to the temperature rise in the local environment. The continuous visible probe beam is used to detect the thermal effect. (b) Noise Power Spectral Density (PSD) along different axes. (c) Noise spectral variation in MIP. Noise in oil is imaged in the fingerprint region. (d) Hyperspectral MIP image of fungal cells in the fingerprint region and the SPEND result at 952 cm$^{-1}$, 1070 cm$^{-1}$, and 1162 cm$^{-1}$. (e) Temporal color-coding result. (f) Zoom-in image of three-layer fungal wall structures. I: inner; M: middle; O: outer. (g) MIP spectrum of each layer. Raw data was published[39].

# Supplementary Information

## Self-Supervised Elimination of Non-Independent Noise in Hyperspectral Imaging

Guangrui Ding[1,4], Chang Liu[2,4], Jiaze Yin[1,4], Xinyan Teng[3,4], Yuying Tan[2,4], Hongjian He[1,4], Haonan Lin[1,4*], Lei Tian[1,2,4*], Ji-Xin Cheng[1,2,3,4*]

[1] Department of Electrical and Computer Engineering, Boston University, Boston, MA, USA, 02215

[2] Department of Biomedical Engineering, Boston University, Boston, MA, USA, 02215

[3] Department of Chemistry, Boston University, Boston, MA, USA, 02215

[4] Photonics Center, Boston University, Boston, MA, USA, 02215

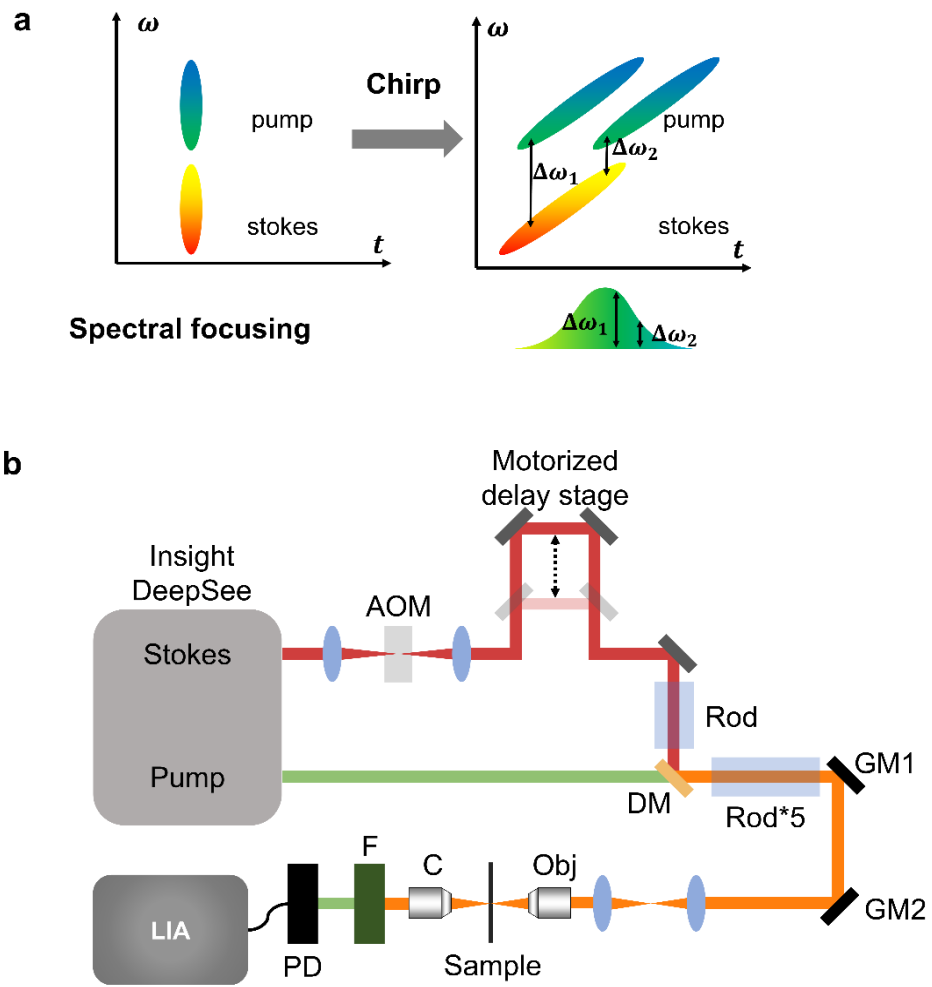

**Figure S1. Hyperspectral Stimulated Raman Scatter setup.** (a) Spectral focusing. Two femtosecond laser pulses are chirped to picosecond and delayed in the time domain. The wavenumber difference of two pulses corresponds to the vibrational state of chemical bonds. (b) Lab-build hyperspectral SRS setup. AOM: acoustic optical modulator; DM: dichroic mirror; GM: Galvo mirror; Obj: Objective; C: sample; F: filter; PD: photodiode; LIA: lock-in amplifier.

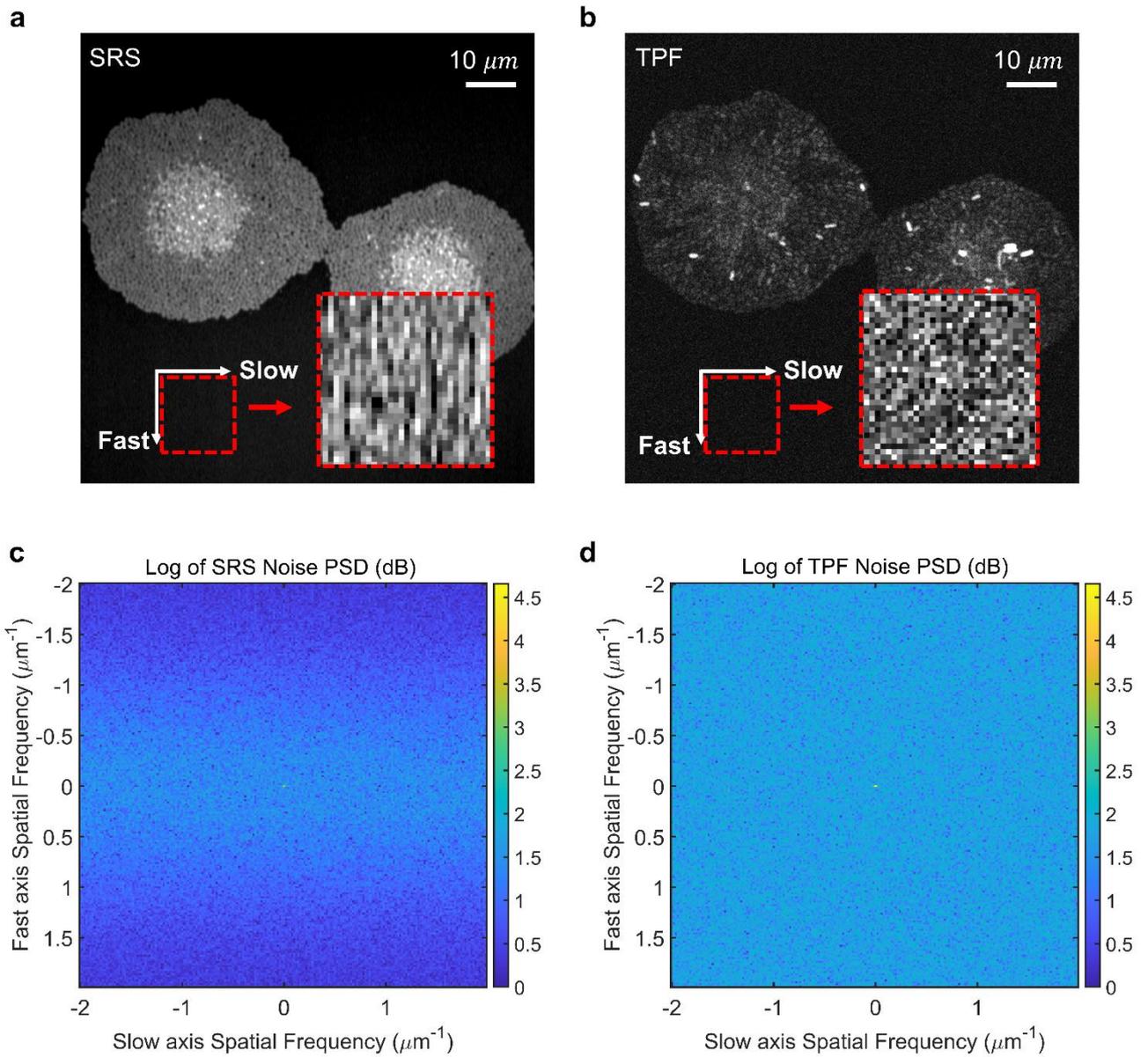

**Figure S2. Fluorescence & SRS spatial noise comparison.** (a) Single color SRS image of OVCAR5. (b) Fluorescence image of OVCAR5. (c) Noise PSD of SRS image. (d) Noise PSD of fluorescence imaging.

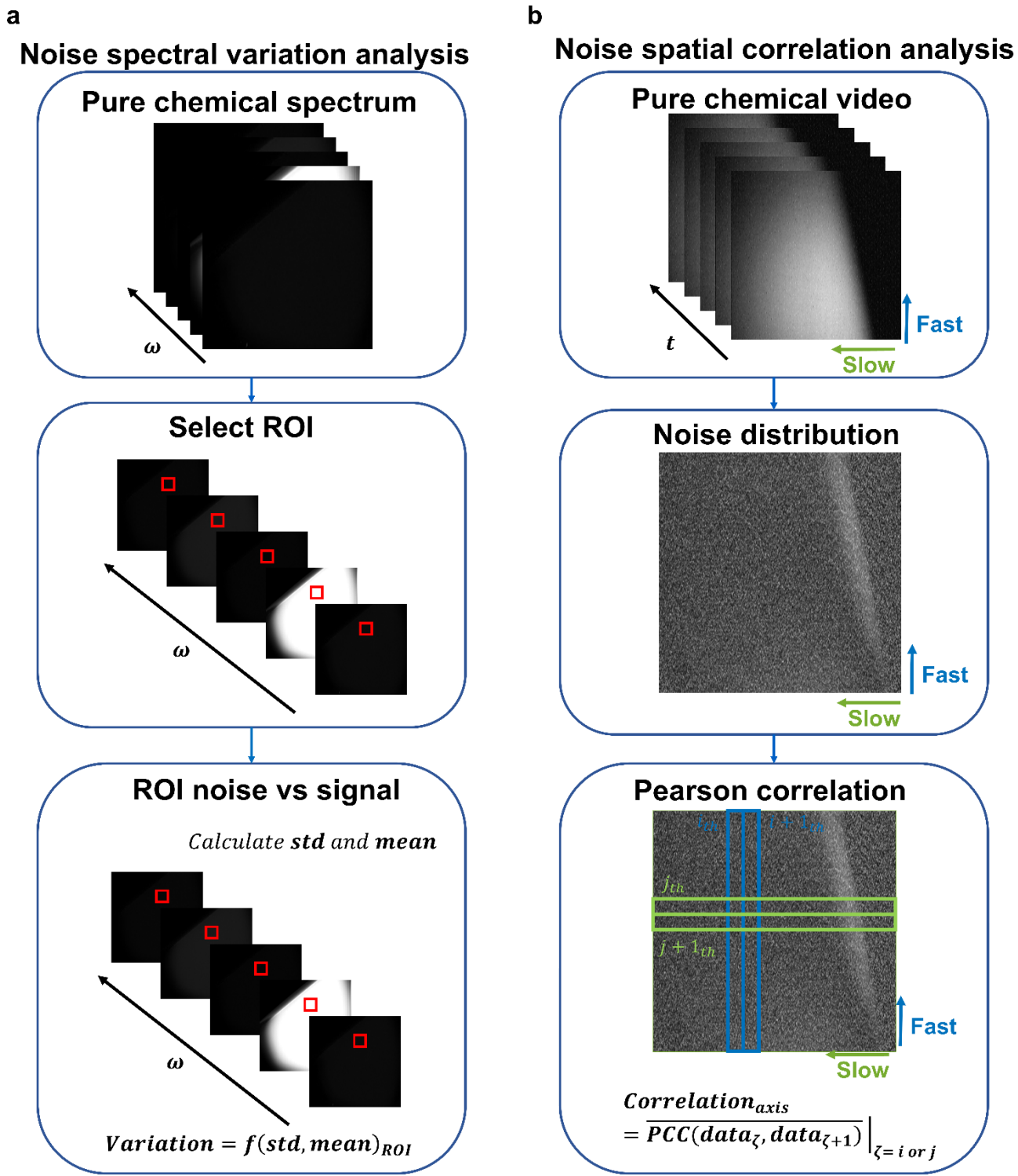

**Figure S3. Noise analysis workflow.** (a) Spectral variation analysis. (b) Spatial correlation analysis.

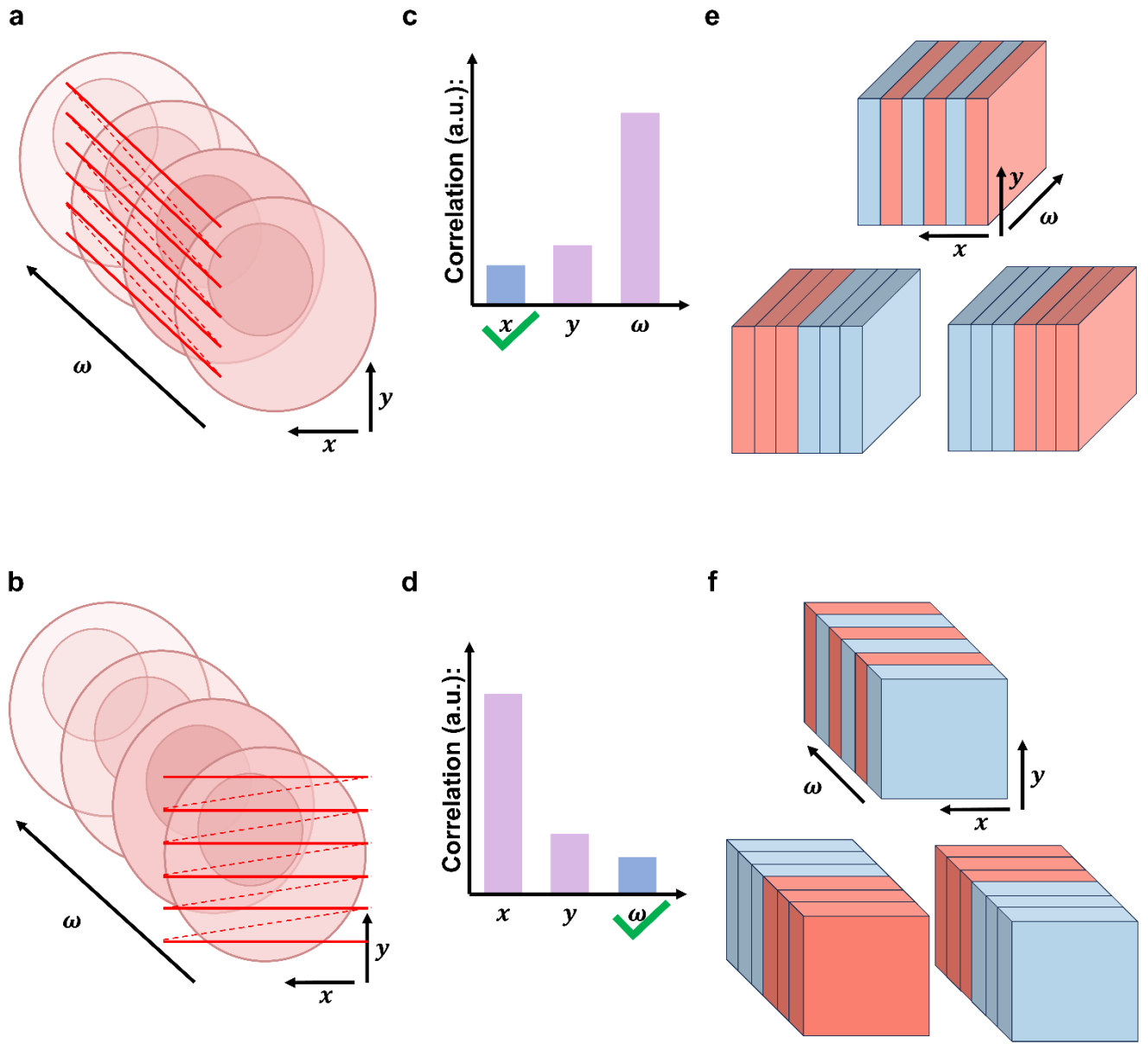

**Figure S4. Different permutation strategies to handle data from different scanning modalities.** It is shown that two scanning modalities lead to two different permutation strategies. (a) $\lambda - x - y$ scanning. (b) $x - y - \lambda$ scanning. (c)-(d) Choose the corresponding axis as the permutation axis according to the fluctuation level. (e)-(f) Permutation based on the chosen axis.

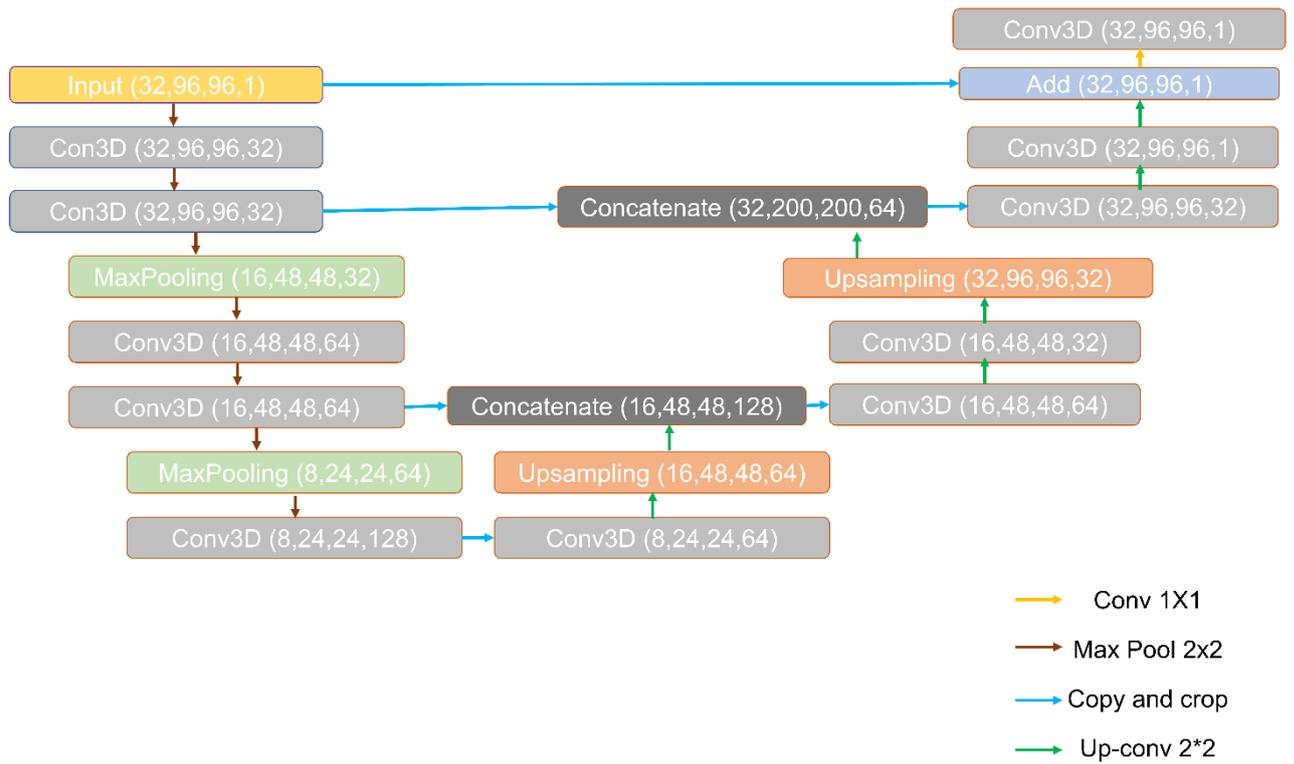

**Figure S5. Network architecture of SPEND.** The whole network is based on a three-layer Unet architecture.

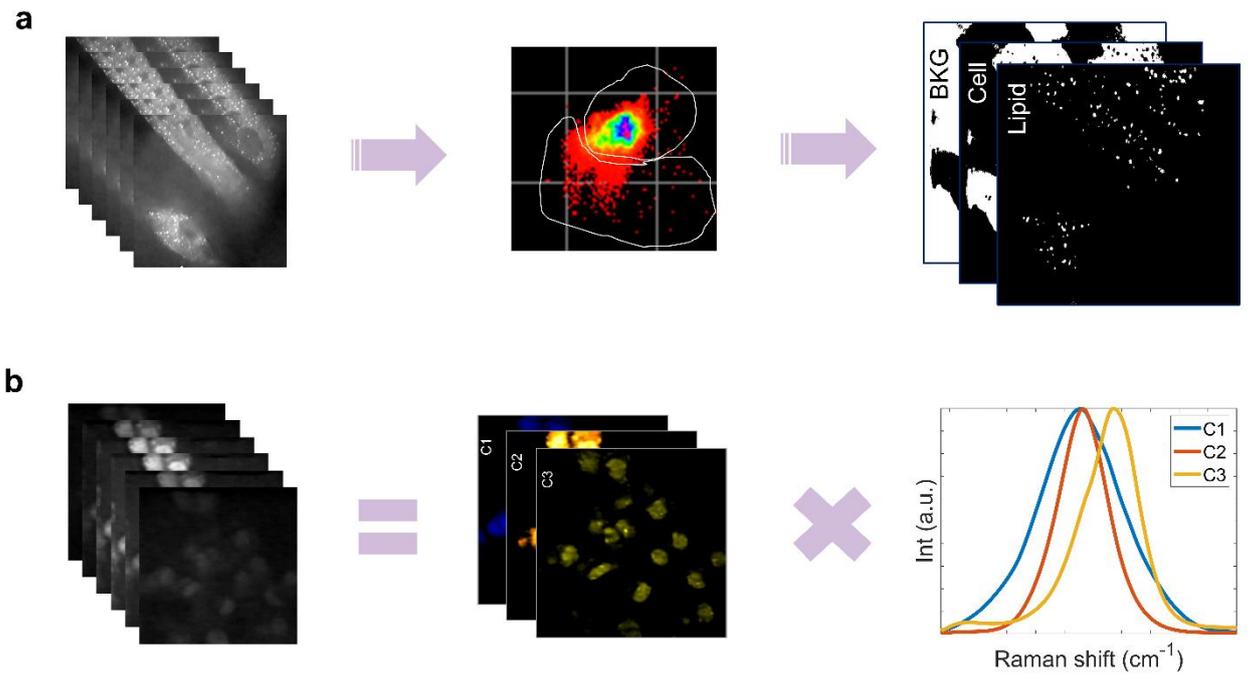

**Figure S6. Chemical unmixing methods.** (a) Unsupervised chemical unmixing methods. No chemical reference is needed. The outputs are segmentation of each component based on the spectral difference. (b) Supervised chemical unmixing methods. Chemical references are needed. The outputs are quantitative chemical concentration maps.

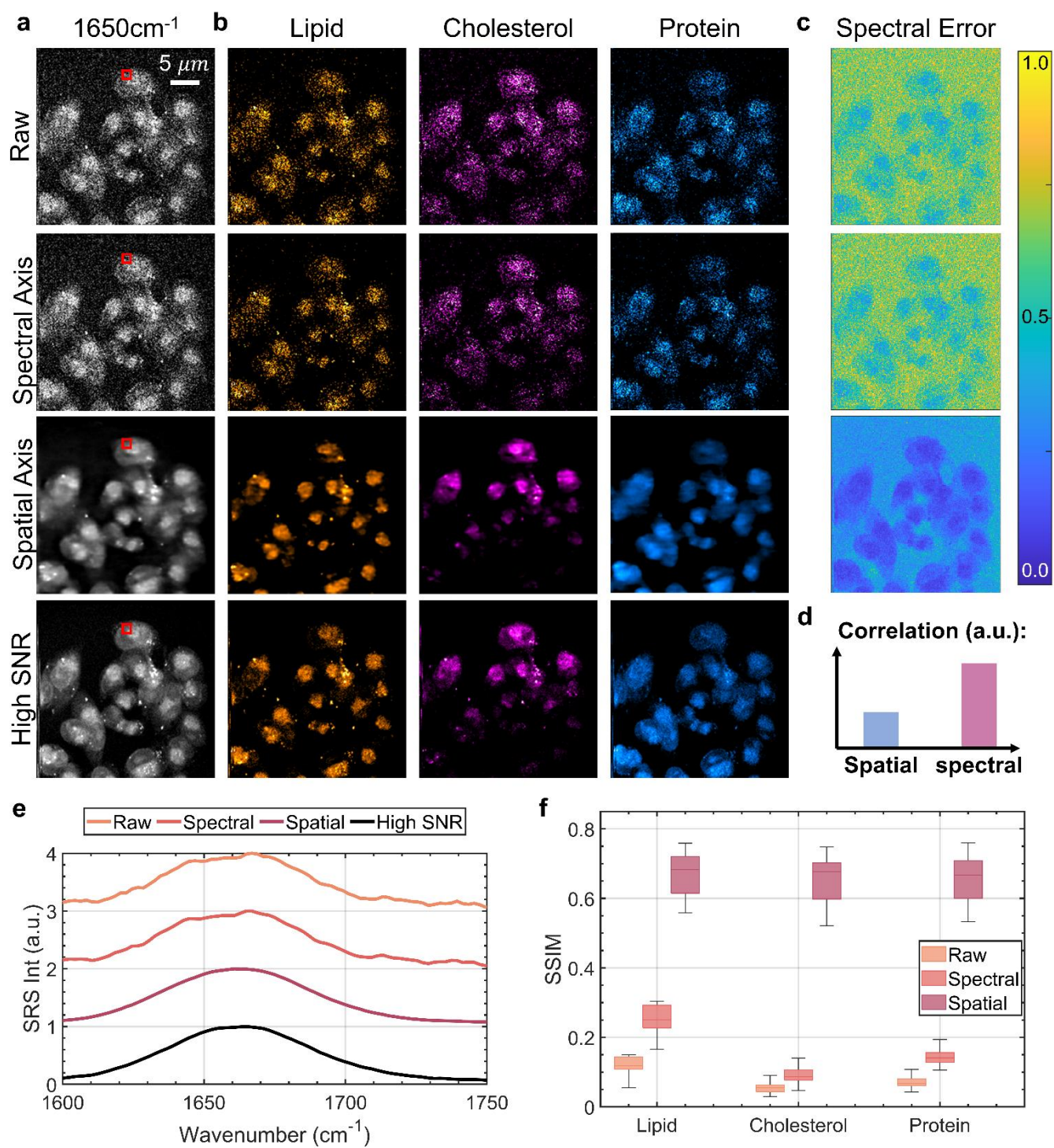

**Figure S7. Performance of permutation along different axes.** (a) Single color SRS (1650cm$^{-1}$) (b) Chemical unmixing map. (c) Spectrum error respectively. (d) Correlation of each axis. (e) Spectrum of ROI. (f) Chemical map SSIM.

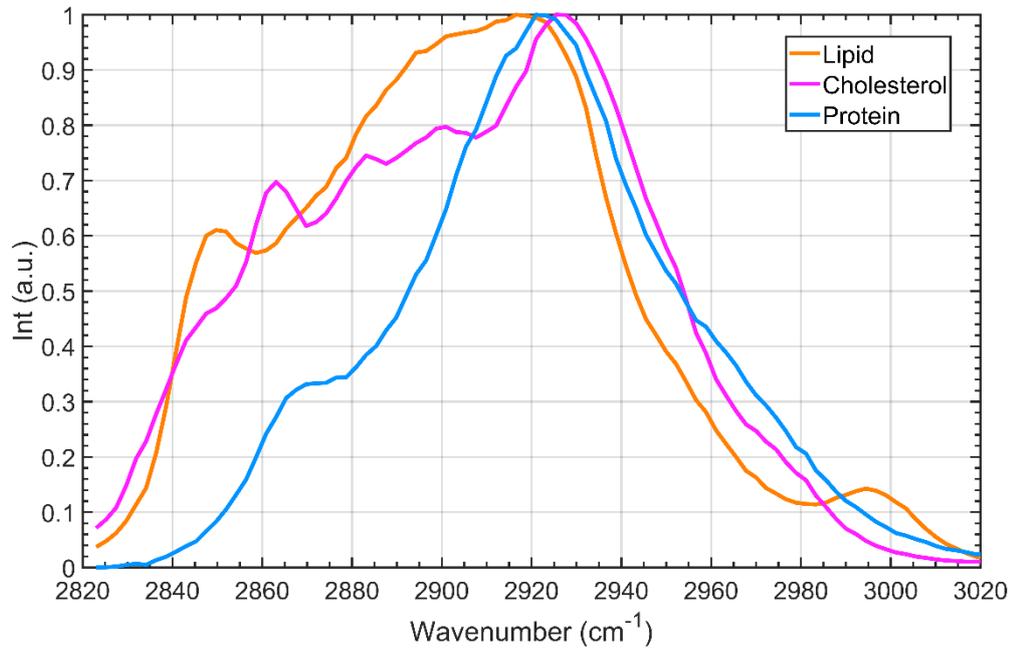

**Figure S8. chemical reference for OVCAR5 spectral unmixing.**

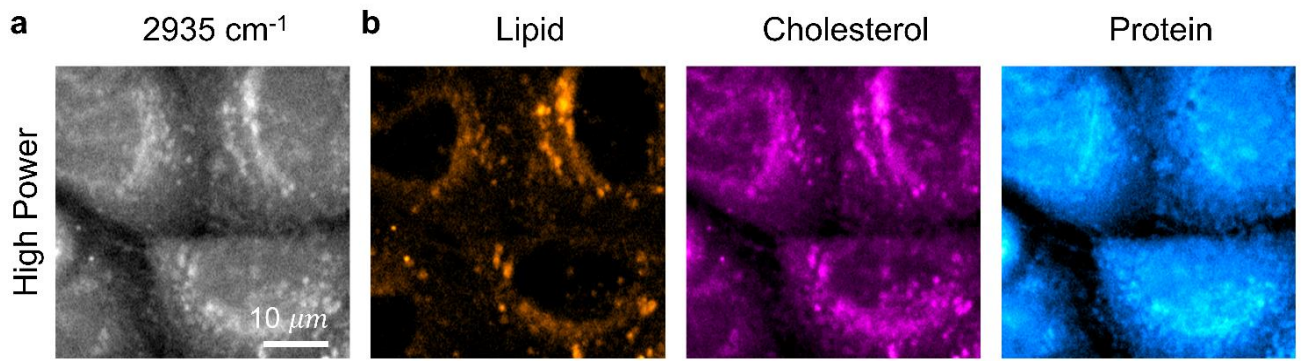

**Figure S9. High Power group of OVCAR5 cell.** (a) Single color SRS image of OVCAR5 cell @2935.2cm$^{-1}$. (b) Chemical unmixing map of OVCAR5 hyperspectral SRS stack.

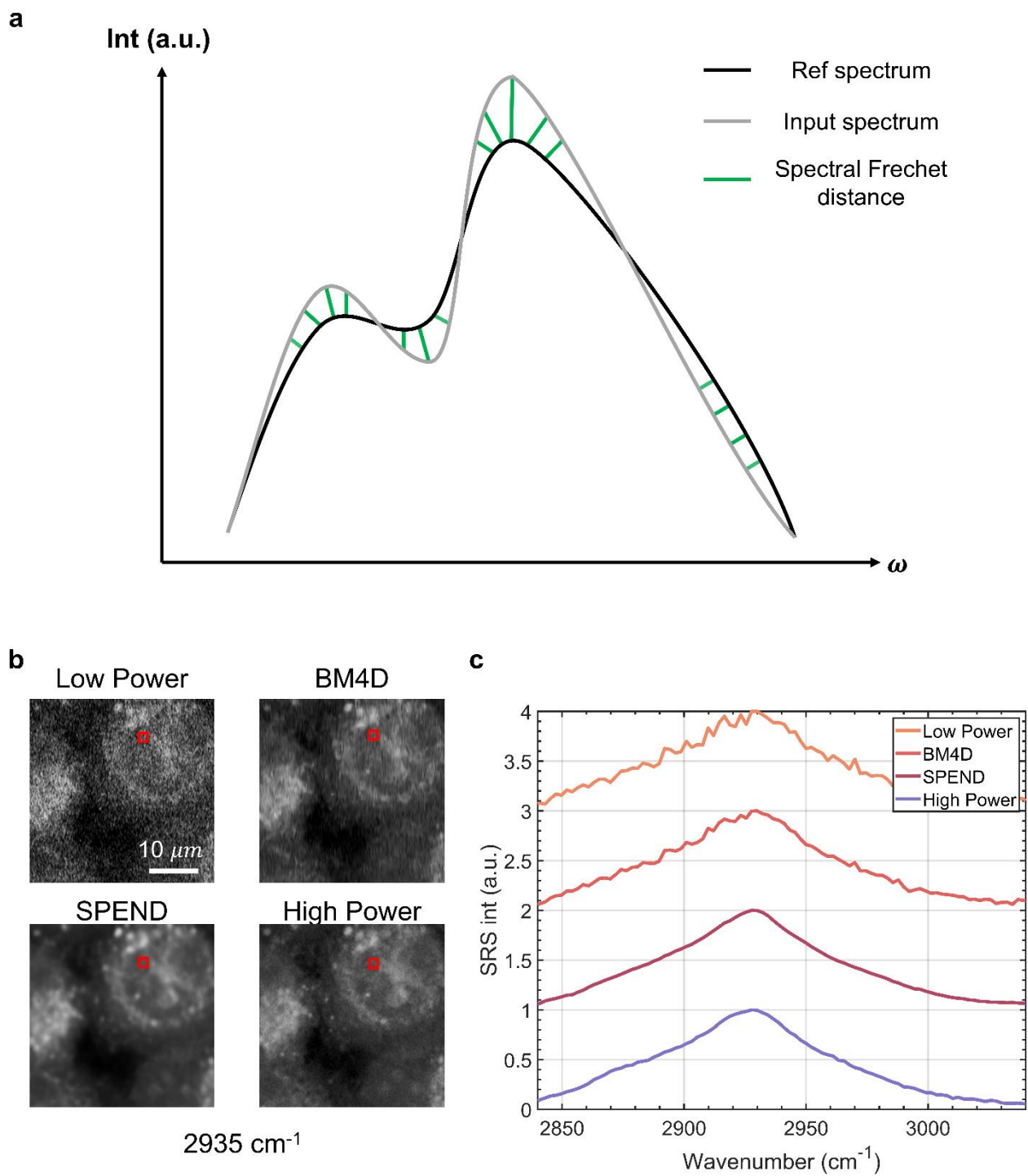

**Figure S10. Spectral distortion calculation diagram.** (a) Fréchet distance diagram. (b)-(c) ROI spectrum plot. Spectrum from ground truth is utilized as the reference. The same ROI spectrums of different groups are input to calculate freshet distance to calculate spectral distortion.

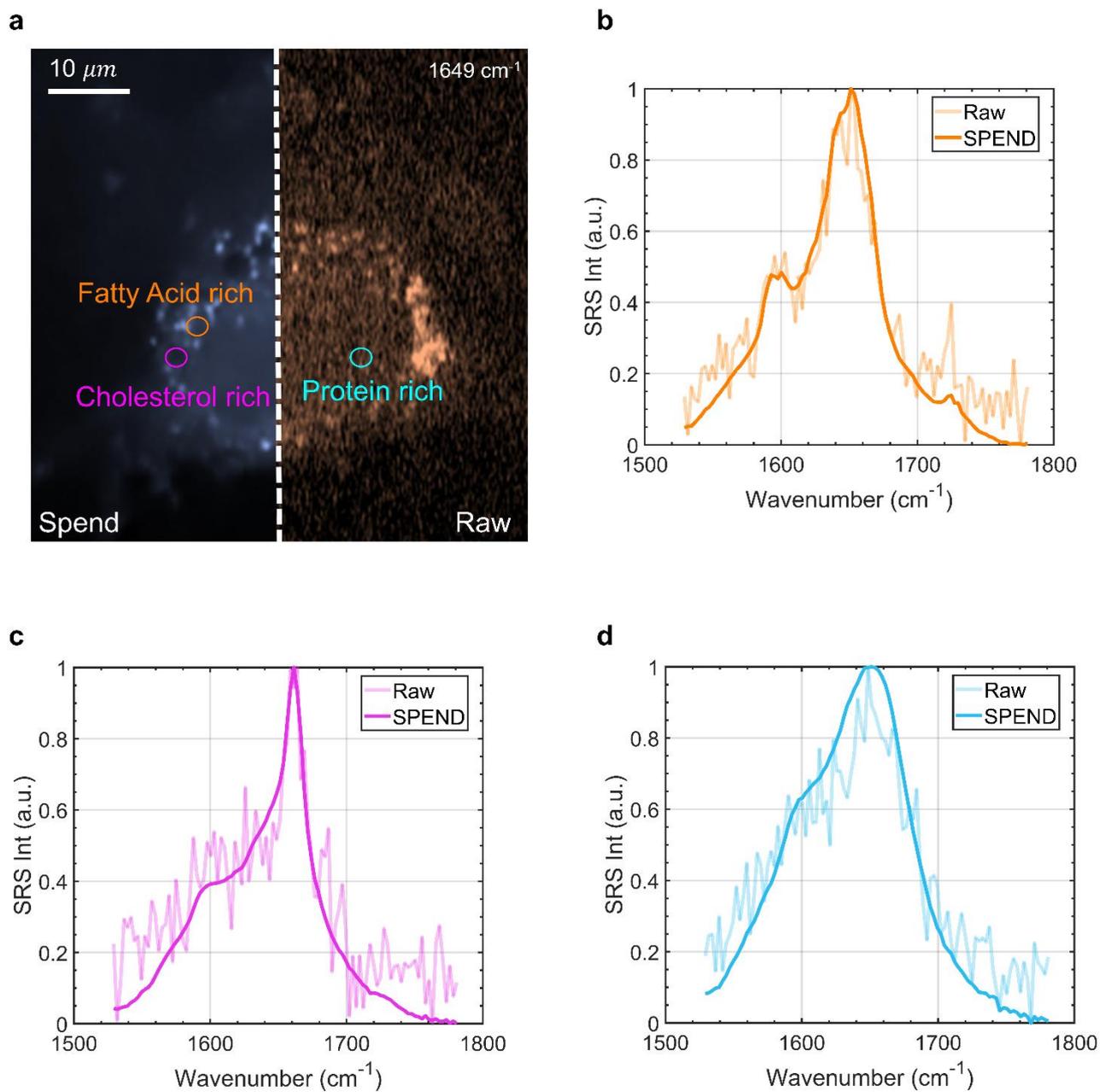

**Figure S11. The spectrum of different chemical rich ROI in the fingerprint region.** (a) single color SRS image of U87 cell @1648.6cm$^{-1}$ (b)-(d) Spectrum of denoising and raw pixel in (b) fatty acid rich, (c)cholesterol rich, and (d) protein rich.

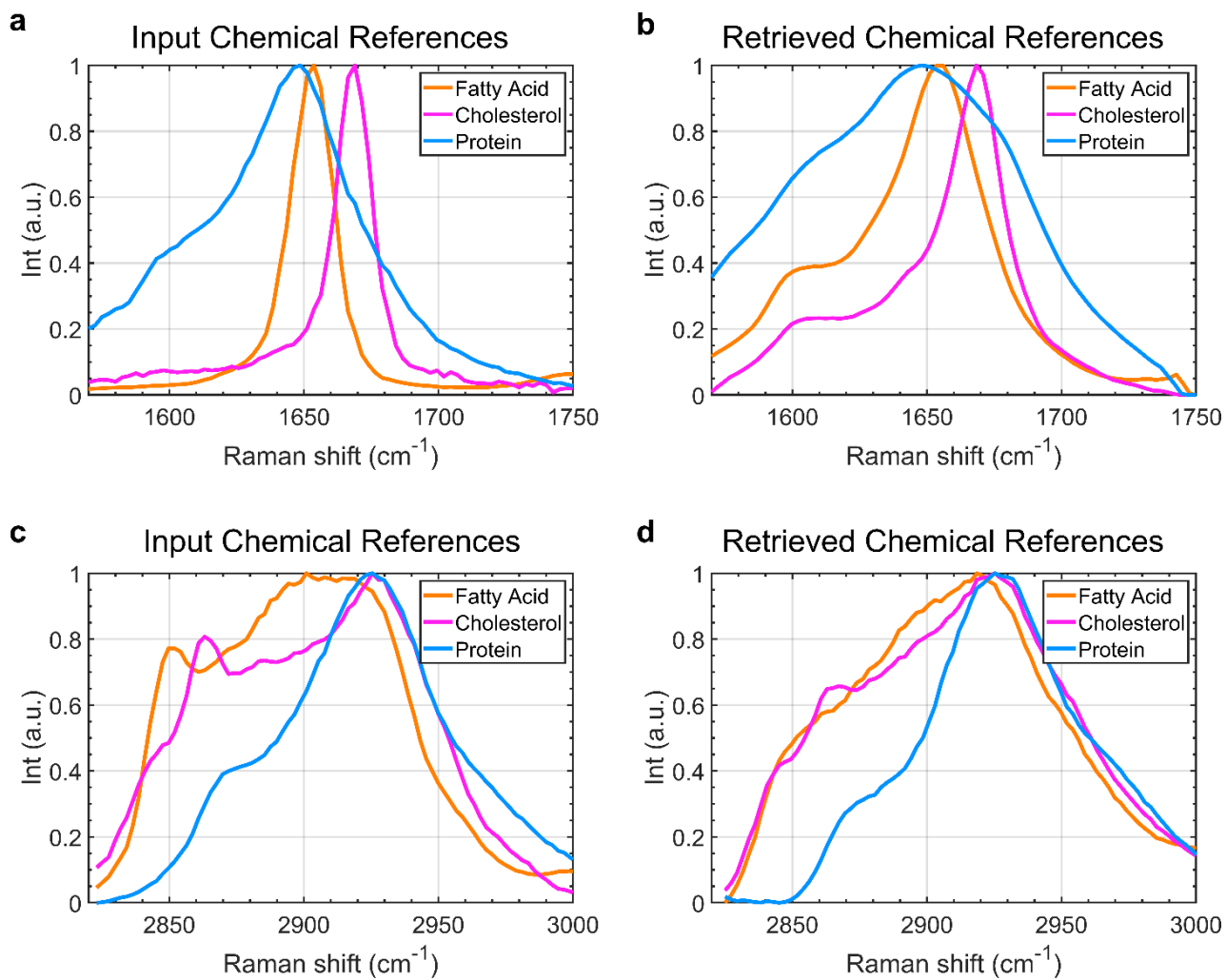

**Figure S12. Comparison of Input chemical references spectrum and MCR retrieved references spectrum.** (a) Input chemical references in the fingerprint region. (b) Retrieved chemical references in the fingerprint region. (c) Input chemical references in the C-H region. (d) Retrieved chemical references in C-H region.

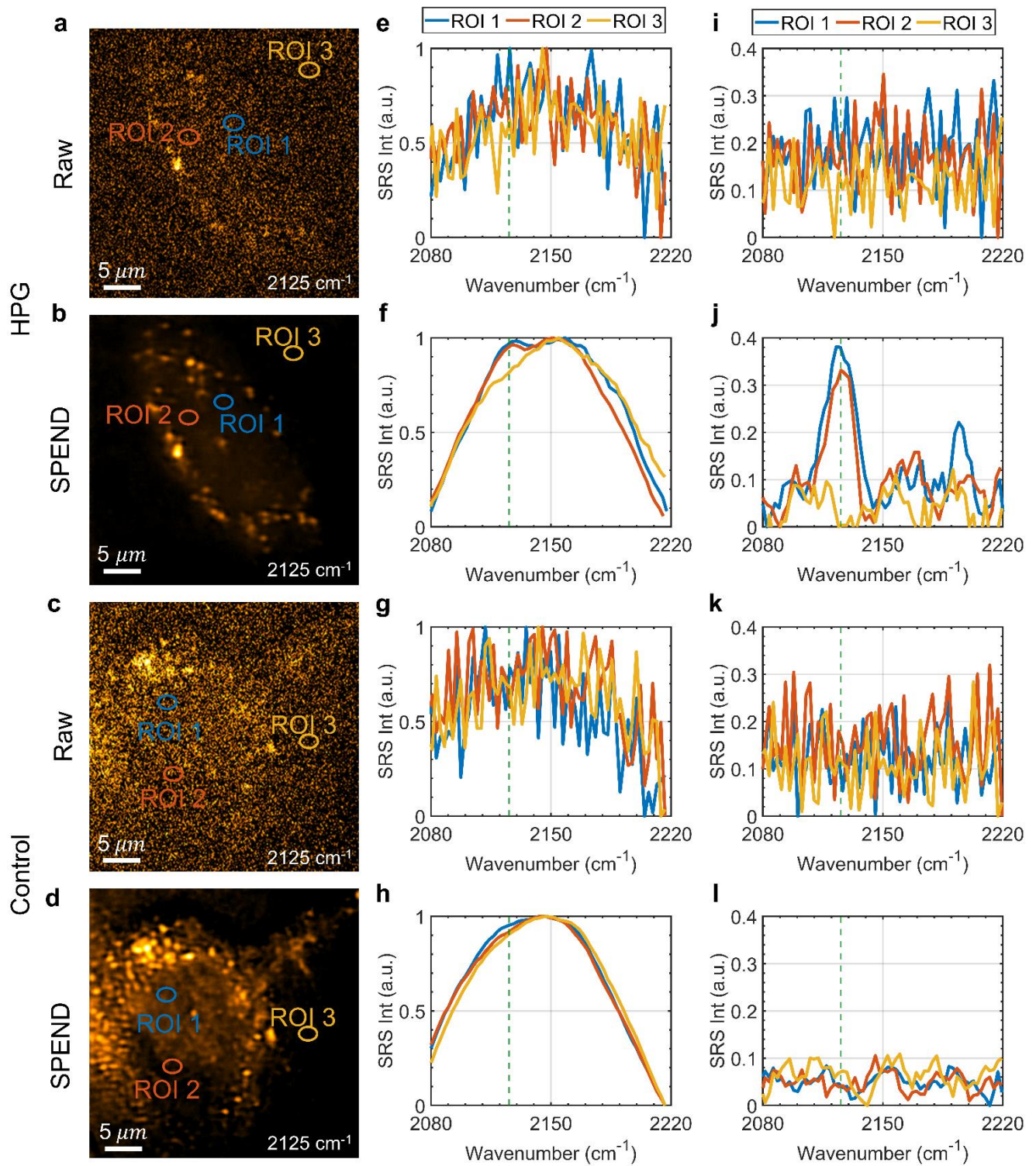

**Figure S13. Spectrum in different ROI.** (a)-(b) Raw and denoising result of single-color SRS of the treated group. @2125cm$^{-1}$. (c)-(d) Raw and denoising result of single-color SRS of control group @2125cm$^{-1}$. (e)-(f) spectrum of ROI of raw and denoising treated group respectively. (g)-(h) spectrum of ROI of raw and denoising control groups respectively. (i)-(j) Normalized spectrum after arPLS baseline correction of HPG treated Group. (k)-(l) Normalized spectrum after arPLS baseline correction of the control group.

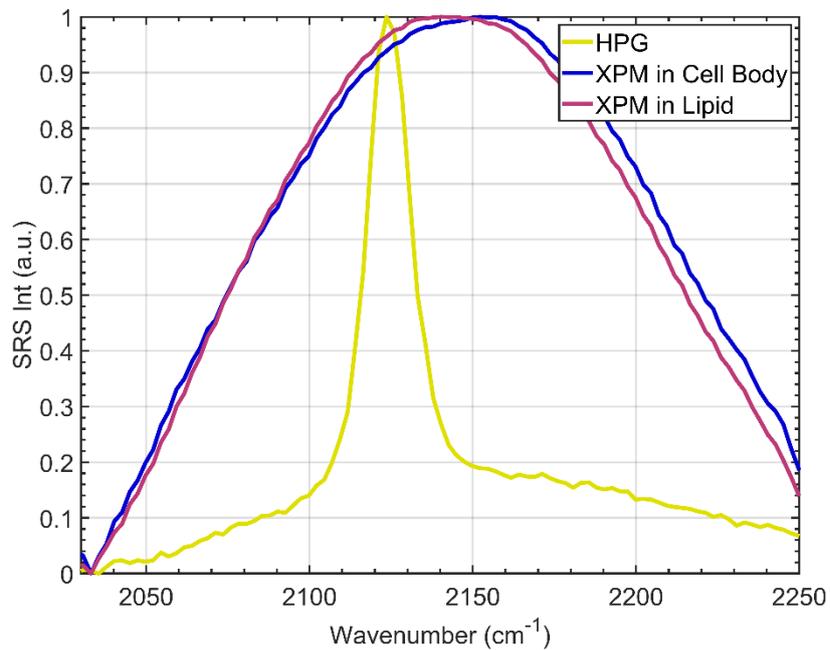

**Figure S14. Reference spectrum generated for lasso unmixing by the retrieval of phasor segmentation result.**

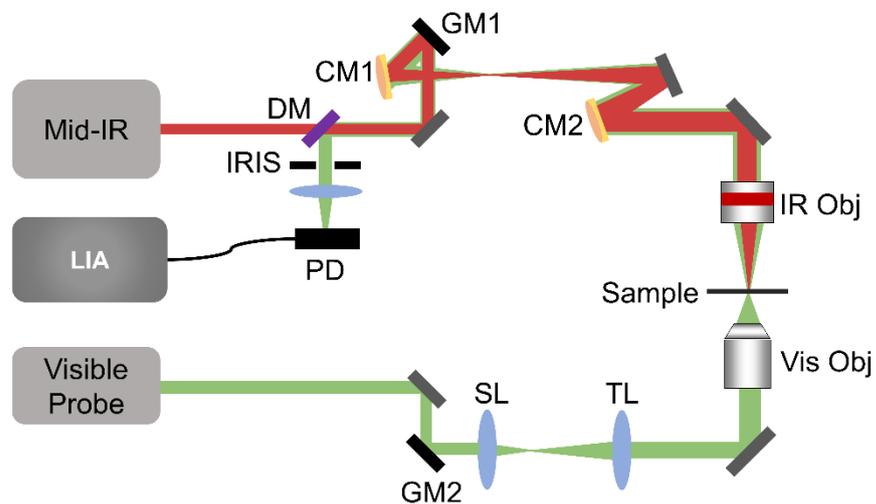

**Figure S15. Mid-infrared photothermal microscopy setup.** DM: Dichroic mirror; CM1: concave mirror f=150 mm; CM2: concave mirror f=250 mm; IR obj: reflective objective lens 40X 0.5NA; Vis Obj: visible objective lens 60X 1.2NA; SL: scan lens f=75 mm; TL: tube lens f=180mm; GV: Galvo mirrors.